\documentclass[12pt]{revtex4-1}
\usepackage{epsfig,bm,dcolumn}
\usepackage{graphicx}
\usepackage{ulem}
\usepackage{epsfig}
\usepackage{latexsym}
\usepackage{bm}
\usepackage{amsmath,amssymb,graphicx,psfrag}
\usepackage[colorlinks,linkcolor={blue},citecolor={blue},urlcolor={blue}]{hyperref}
\usepackage{caption}
\usepackage{xcolor}
\usepackage{graphicx}
\usepackage{dcolumn}
\usepackage{bm}
\usepackage{amsmath}
\usepackage{amssymb}
\usepackage{float}
\usepackage{physics}
\usepackage{comment}

\begin{document}



\title{Periodic drive induced unconventional superconductivity in a half-filled system}

\author{Suryashekhar Kusari}
\affiliation{Theory Division, Saha Institute of Nuclear Physics, 1/AF Bidhannagar, Kolkata 700064, India}
\affiliation{Homi Bhabha National Institute, Training School Complex, Anushaktinagar, Mumbai 400094, India}
\author{Arti Garg}%
 \affiliation{Theory Division, Saha Institute of Nuclear Physics, 1/AF Bidhannagar, Kolkata 700064, India}
\affiliation{Homi Bhabha National Institute, Training School Complex, Anushaktinagar, Mumbai 400094, India}

\begin{abstract}
The non-equilibrium control of electronic properties has emerged as a transformative paradigm for engineering novel quantum phases. The most intriguing example of such a phase is light-induced superconductivity (SC) in non-superconducting materials. However, realizing unconventional SC at commensurate half-filling remains a formidable challenge even in non-equilibrium, as the regime is typically dominated by the robust stability of the antiferromagnetic (AFM) Mott insulating (MI) state. Here, we provide a novel non-equilibrium route to realize unconventional d-wave SC in a half-filled system through Floquet engineering. We analyze the periodically driven Fermi-Hubbard model on a bipartite lattice and demonstrate that a high-frequency drive can transform a weakly interacting insulator into a regime of strong correlations by the drive-induced renormalization of nearest-neighbor hopping. Furthermore, the drive induces staggered  higher range hoppings that can frustrate the AFM order while simultaneously generate staggered potential that lifts the kinetic constraints inherent to the half-filled system, fostering the charge dynamics required to stabilize d-wave pairing against the competing AFM state. The resulting SC phase is protected by high-frequency prethermalization, maintaining stability over timescales exponentially large in the drive frequency. This protocol circumvents the need for chemical doping, offering a 'disorder-free' alternative for realizing unconventional pairing with direct applications in optimizing the performance of superconducting quantum computers, qubit arrays and other upcoming quantum technologies.

\end{abstract}

\maketitle

\section{Introduction} 
Floquet engineering has revolutionized the design of quantum states of matter by utilizing time-periodic perturbations to transcend the inherent limitations of static equilibrium. By employing time-periodic perturbations to coherently drive electronic degrees of freedom, this framework enables the synthesis of exotic phases that lack any static equilibrium counterpart. This approach has already facilitated seminal breakthroughs, including the observation of the light-induced magnetic order~\cite{Mag_order}, light induced charge density and spin-density wave~\cite{AFMorder,charge_order}, topological phases like anomalous Hall effect in monolayer graphene ~\cite{graphene_AQH} and 3D topological insulators~\cite{AQHE}, light-induced many-body localization~\cite{MBL_light} and discrete-time crystals~\cite{DTC1, DTC2, DTC3,Aarya} etc.
Perhaps the most remarkable phenomenon within this field is the emergence of light-induced superconductivity (SC) by exciting samples with lasers in the non-superconducting phase. Superconductor-like behavior has been experimentally observed by photo-excitation in many systems including cuprates~\cite{light_sc_cuprates,light_cuprates2,Cavalleri_PRX} , fullerides~\cite{fullerene1,fulleren2,fulleren_sc} and organic molecular crystals~\cite{organic}. While these experiments suggest that optical driving can enhance or initiate SC, a robust theoretical framework for inducing unconventional pairing in otherwise insulating systems by driving remains a theoretical challenge. It is hence highly desirable to investigate the dynamical onset of light-induced SC theoretically. 

With this motivation, in this work, we present a novel non-equilibrium route to engineer unconventional SC at half-filling by periodically driving a weakly interacting system. By investigating the periodically driven Hubbard model and its variants on a bipartite lattice at half-filling, we demonstrate that periodic modulation can fundamentally restructure the e-e correlations to favor a spin-exchange mediated unconventional SC phase over a wide range of drive parameters. Our findings provide a new pathway for utilizing light to foster high-temperature unconventional SC in a half-filled system. This is especially interesting because in most unconventional superconductors, the SC is typically accessed by tuning the parent compound away from commensurate filling ~\cite{Bednorz,pnictide_expt,Lee,Pnictides,MTBLG,MTBLG2} through chemical doping (barring a few~\cite{MTBLG,MTBLG2}) which inevitably introduces quenched disorder~\cite{Pan,Mcelroy,Garg}. Realizing unconventional SC at commensurate filling remains a challenge in condensed matter physics. In this work, we demonstrate that Floquet engineering offers a pristine, non-equilibrium alternative to achieve this, bypassing the deleterious effects of chemical disorder.
\begin{figure}
  \includegraphics[width=4.75in,angle=0]{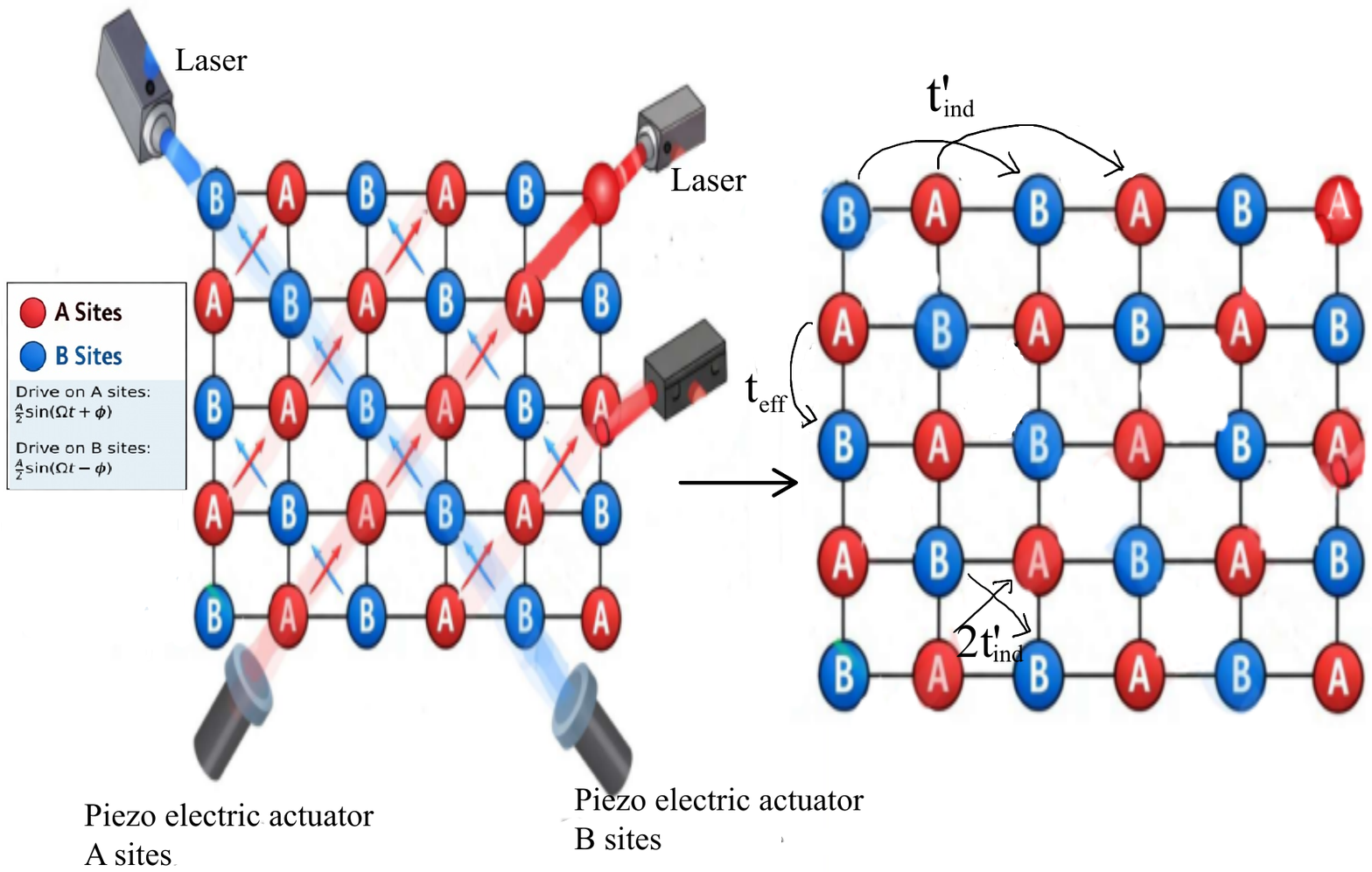}
 \includegraphics[width=4.0in,angle=0]{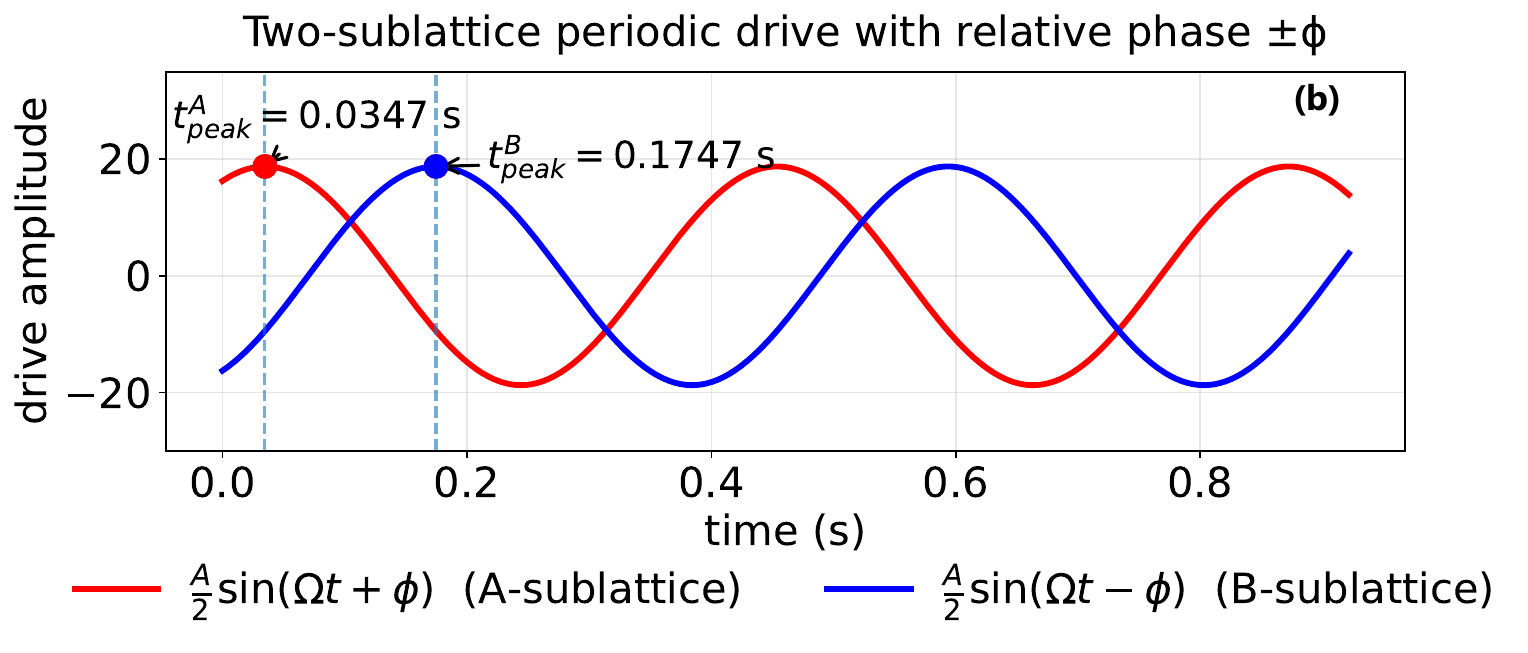} 
 
    \caption{Panel[a]: A possible implementation of the periodic drive explored in this work in ultracold experiments. The right panel shows the drive induced 2nd and 3rd neighbour hoppings on 2-dimentional square lattice. Panel [b] shows how time lag between the drive of A and B sublattices can induce the required phase difference $\Phi$.}
    \label{Fig1}
    \end{figure}

A central challenge in the study of periodically driven many-body systems is the propensity for the drive to act as an infinite energy reservoir, eventually leading to a featureless, infinite-temperature state~\cite{LDM_PRE,Alessio2014}. This process, often referred to as "Floquet thermalization," typically occurs when the driving frequency is resonant with the characteristic energy scales of the undriven Hamiltonian, driving the system toward a state of maximum entropy.
Despite this tendency toward heating, stability can be recovered in limits where the drive frequency and amplitude outpaces the system’s natural relaxation rates, granting access to long-lived non-equilibrium states. In the high-frequency regime, $\Omega\gg W$ where $W$ is the local bandwidth of the system, any generic short range interacting system  enters a long-lived prethermal state in which  energy absorption is suppressed 
resulting in heating timescales exponentially large in the driving frequency: $t_{th} \ge O(\exp{\Omega/W})$~\cite{Dima_Pretherm_PRL, Mori_Kuwahara_Saito_Prethermal_PRL, Kuwahara_Mori_Saito, Dima_Floquet_Prethermalization,Prethermal_Without_T, Mori_2022, Ho_Mori_Abanin_Pretherm_Rev}. 
This exponential suppression of heating rate ensures a wide window of quasi-stability, during which the system is governed by a static effective Floquet Hamiltonian. Parallel to this, the strong-amplitude regime offers stability through the renormalization of kinetic energy scales. 
This prethermal window is of critical experimental relevance, as it allows for the observation and manipulation of non-equilibrium phases—such as the unconventional SC proposed here over experimentally relevant timescales. 

There have been earlier studies on the periodically driven Hubbard model to explore the possibility of SC~\cite{Coulthard,Chiral_SC,TGA,Zeng_sc} but none of these works have investigated the possibility of unconventional SC at half-filling , which is the main focus of our work. Here, we show that periodically driving a weakly interacting Hubbard model and its variants (one having a staggered potential) on a bipartite lattice at half-filling, can induce d-wave SC as well as antiferromagnetic (AFM) Mott insulating (MI) phase. In this high-frequency, high-amplitude regime, the bare nearest-neighbor hopping is exponentially suppressed to an effective value $t_{eff}$, which drives the system into a strongly correlated regime ($U/t_{eff}\gg1$). Beyond simple renormalization, the drive induces a staggered potential and staggered higher-order hopping terms that can frustrate the AFM order, preventing the system from collapsing into a MI state for a certain range of drive parameters. Crucially, this protocol resolves the doping-disorder paradox through a mechanism of dynamic density imbalance. Although the global system remains at half-filling, the induced (or enhanced) staggered potential creates a local environment where the two sublattices are effectively electron- and hole-doped. This spatial redistribution of charge facilitates the carrier dynamics necessary for d-wave SC, all while maintaining the integrity of the commensurate parent compound. The unconventional superconducting phase, proposed here, is central to quantum technologies due to its essential role in superconducting qubit arrays and quantum computer~\cite{qubits}. 


\noindent
\textbf{Periodically Driven Hubbard Model} \\
We focus on the generalized Fermi-Hubbard model on a bipartite lattice in the presence of a sublattice dependent periodic drive. The model is
described by the following time-dependent Hamiltonian.  
\[H(t) =H_{Hub}+H_{drive}(t)\] 
\begin{equation}
\begin{split}
 H_{Hub}=-t_0\sum_{\langle ij\rangle,\sigma}c^\dagger_{iA\sigma}c_{jB\sigma}+ h.c.\
+U\sum_{i}n_{i\uparrow}n_{i\downarrow}\\-\mu\sum_i n_i +V_0\sum_{i}n_{iA\sigma}-V_0\sum_{i}n_{jB\sigma},
 \\
 H_{drive}(t)=\frac{A}{2}sin(\Omega t+\Phi)\sum_{j\in A}\hat{n}_{j}+\frac{A}{2}sin(\Omega t-\Phi)\sum_{j\in B}\hat{n}_{j}
\end{split}
\label{hamil}
\end{equation}

Here, $t_0$ is the nearest neighbour hopping, $U$ is the onsite repulsion between electrons and $\mu$ is the chemical potential that fixes the average density of the system. {\it {We will study the system with and without the staggered potential $V_0$}}. $H_{drive}(t)$ describes a time-periodic single particle Hamiltonian with driving frequency $\Omega$ and amplitude $A$ which are the same for both sublattices, and a phase difference $2\Phi$ between two sublattices. 
\begin{figure}
 \includegraphics[width=2.85in,angle=0]{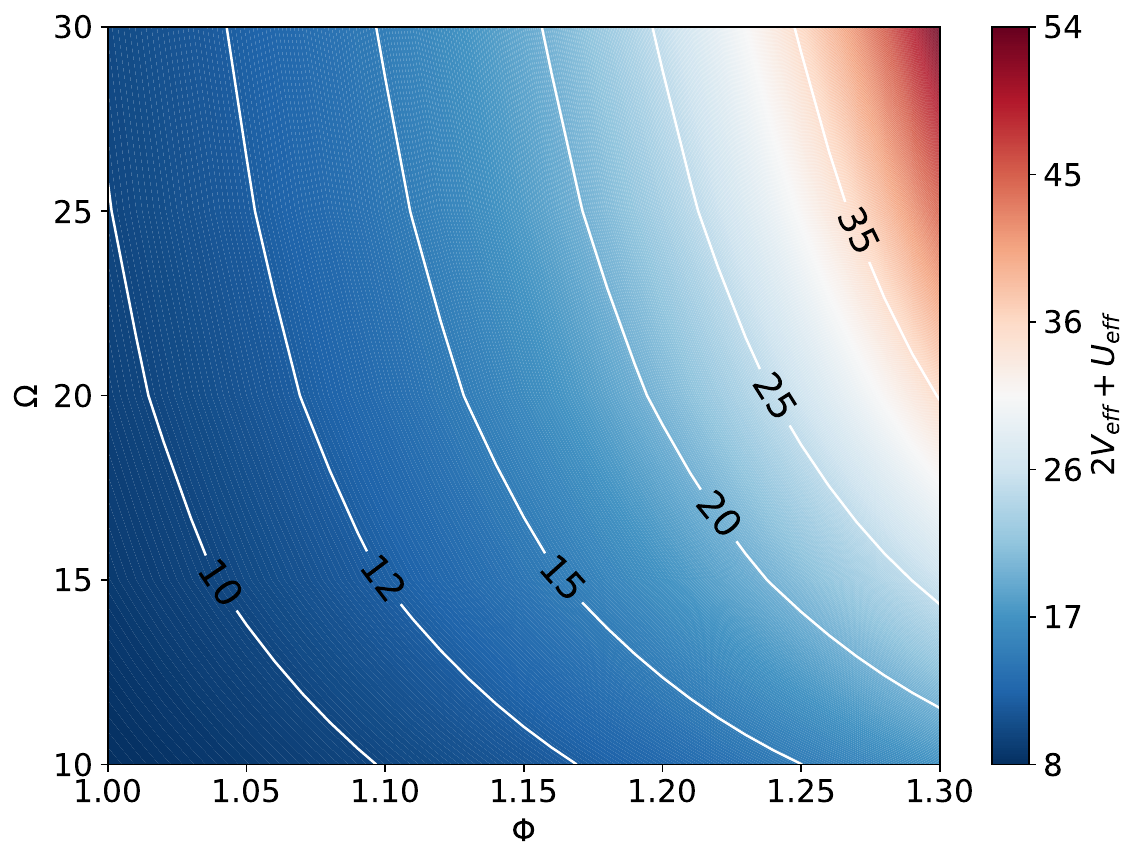}
   \includegraphics[width=2.85in,angle=0]{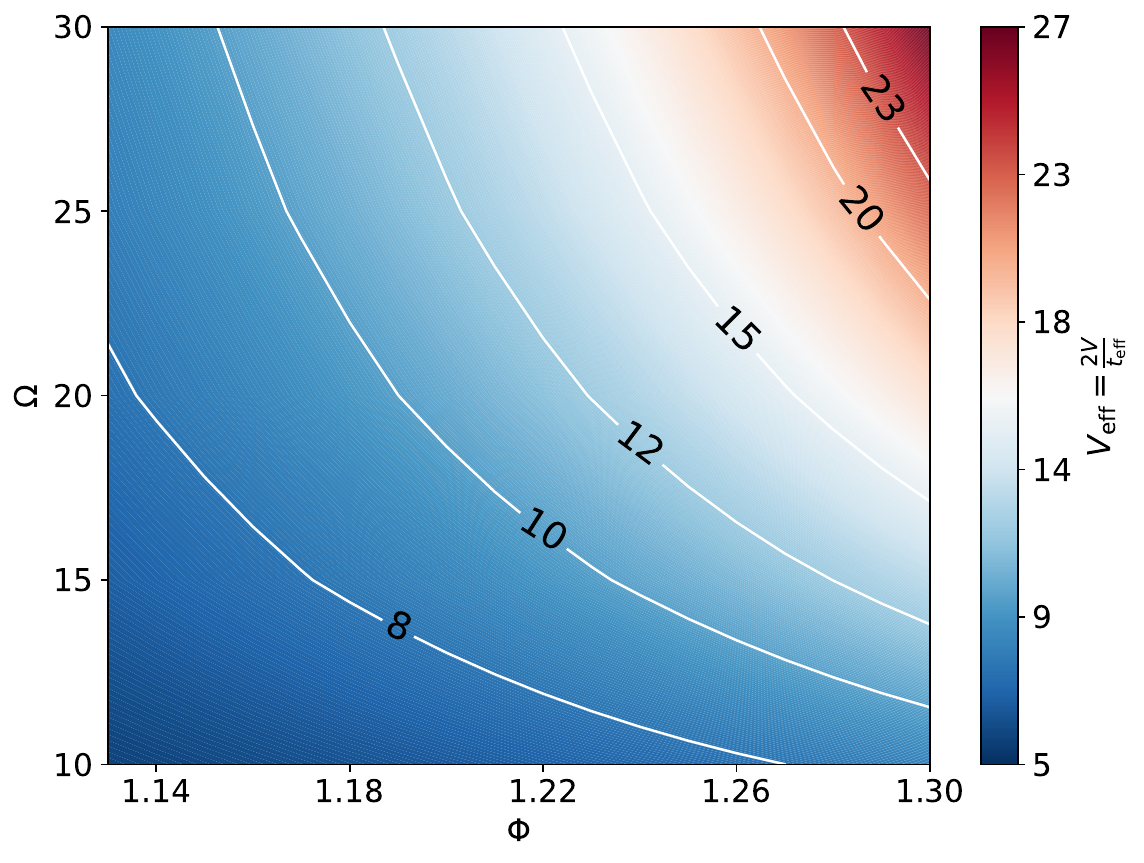} 
   \includegraphics[width=2.85in,angle=0]{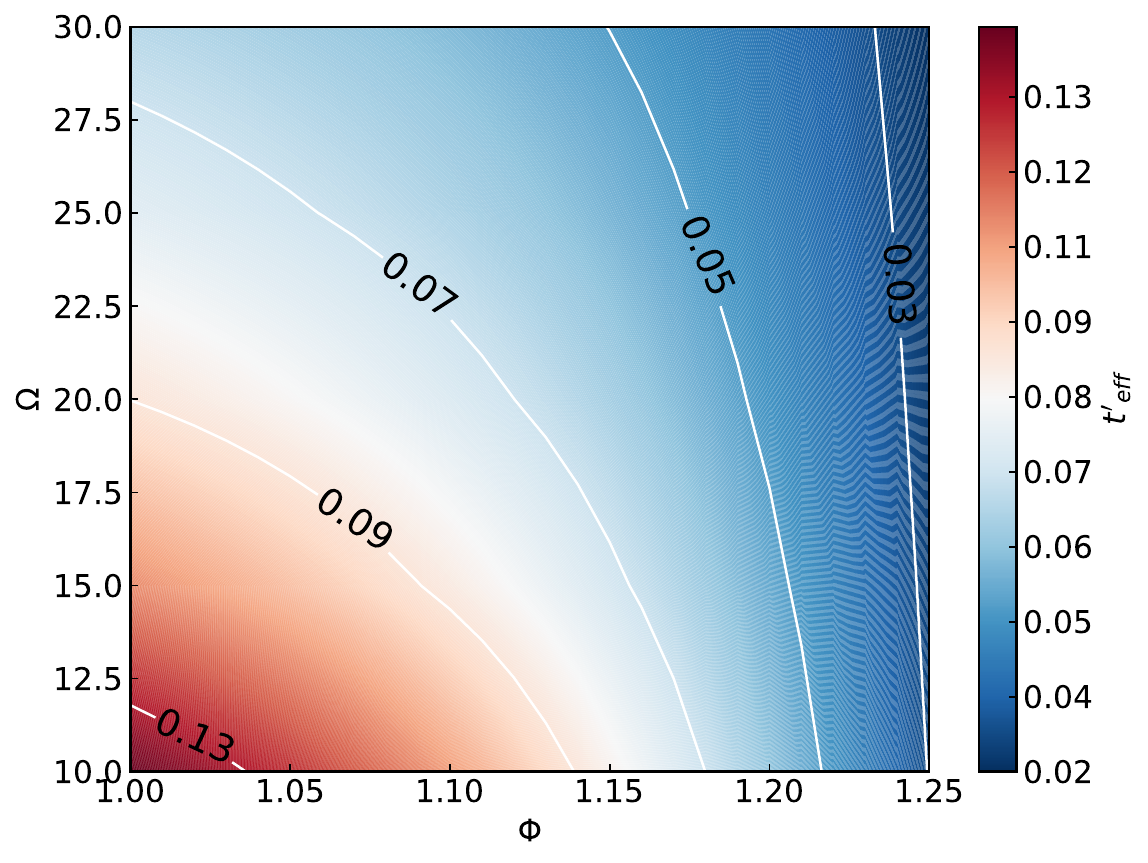} 
     \includegraphics[width=2.85in,angle=0]{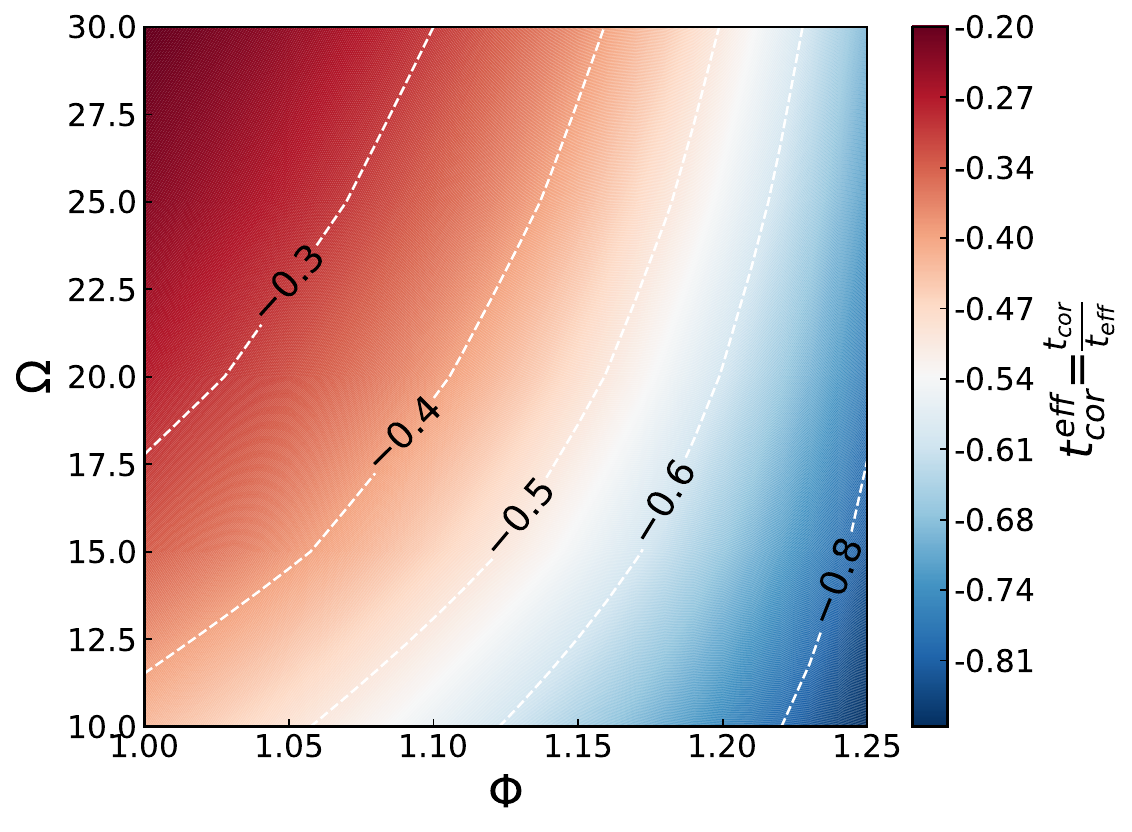} 
    \caption{Effective couplings in periodically driven Hubbard model in $\Omega-\Phi$ plane for the drive with amplitude $A=2.5\Omega$. Here $U_{eff}=U/t_{eff}$ is the effective Hubbard interaction, $V_{eff}=2V/t_{eff}$ is the effective staggered potential between two sublattices and $\pm t^\prime_{eff}=\pm t^\prime_{ind}/t_{eff}$ is staggered higher range intra-sublattice hopping. The right bottom panel shows the effective correlated hopping $t^{eff}_{cor}$ on nearest neighbour bonds. The data shows are for $U_0=1$ and $V_0=0.5$ in units of $t_0$.} 
    \label{Fig2}
          \end{figure}

In the limit of high frequency and large amplitude for the periodic drive, the stroboscopic Floquet Hamiltonian $H_F=H_F^{[0]}+H_F^{[1]}$ obtained using the rotating wave approximation and high frequency Magnus expansion up to order $\frac{1}{\Omega}$ is as follows:
\begin{widetext}
\begin{equation}
\begin{split}
    H_F=-t_{eff}\sum_{\langle ij \rangle, \sigma} c_{i,\sigma}^\dagger c_{j,\sigma} + h.c + U \sum_j n_{j \uparrow} n_{j \downarrow}-\mu\sum_{i\sigma} n_{i\sigma}, \\
    +V\sum_{\langle ij\rangle,\sigma}  \big( n_{i \sigma}^A - n_{j, \sigma}^B \big)+2t'_{ind}\sum_{(ij),\sigma}\big( c_{iA,\sigma}^\dagger c_{jA,\sigma}
- c_{iB, \sigma}^\dagger c_{jB, \sigma} \big) \\
+t_{cor}\sum_\sigma \big(n_{i \sigma}^A -n_{j, \sigma}^B \big)c_{iA \bar{\sigma}}^\dagger c_{jB, \bar{\sigma}} + h.c.
+t'_{ind}\sum_{((ij))} c_{iA,\sigma}^\dagger (c_{jA,\sigma} -c_{iB,\sigma}^\dagger c_{jB,\sigma})
\end{split}    
\label{HF1}
\end{equation}
\end{widetext}
Here, $t_{eff} = t_0 \mathcal{J}_0(\frac{A}{\Omega}sin(\Phi))+t_{ind}$ where $\mathcal{J}_0$ is the zeroth-order Bessel function of the first kind. For the range of parameters where $\mathcal{J}_0$ hits zero, the effective hopping of nearest neighbors is suppressed, resulting in the well-known dynamical localization of the system~\cite{Dunlap, Holthaus, Eckardt, Tsuji}. The drive also induces staggered second and third neighbor hoppings of strength $2t^\prime_{ind}$ and $t^\prime_{ind}$, respectively, providing dynamics within each sublattice along with a staggered potential $V_{ind}$ between sublattices A and B which renormalizes the bare staggered potential to give $V=V_0+V_{ind}$. Another interesting term induced by the drive is a spin-dependent correlated hopping $t_{cor}$ which prevents hopping of holon and doublon across the two sublattices. Note that even if one starts with a pure Hubbard model without any staggered potential, the drive induces a staggered potential $V_{ind}$. The details are provided in the Supplemental Material (SM)~\cite{SM}).

The proposed Floquet protocol is ideally suited for implementation in ultracold atomic systems, where optical lattice geometries can be engineered with sub-wavelength precision~\cite{Bloch2008,Bloch_Fermi_Hubbard,Esslinger1,Esslinger2}. Fig.~\ref{Fig1} shows a schematic diagram for the implementation of the periodic drive in cold atom experiments. 
A bipartite lattice is realized by intersecting two independent pairs of counter-propagating laser beams oriented at $\pi/4$ and $3\pi/4$ degree relative to the x-axis. This configuration creates two interpenetrating square lattices, allowing for the independent manipulation of sublattice degrees of freedom.
Periodic modulation is achieved by mounting the retro-reflecting mirrors on piezoelectric actuators to induce high-frequency longitudinal displacements of the standing waves. In the co-moving frame, this translates to time-periodic inertial forces that drive the system into the high-frequency, large-amplitude regime.To establish the necessary phase difference $\Phi$ between the sublattice drives, we propose a controlled temporal offset in the drive initiation. For a driving frequency $\Omega$, the required phase shift is mapped to a time lag:
$\Delta t=2\Phi/\Omega$.  
Though ultracold atoms provide an ideal architecture for realizing the proposed periodically driven system, we believe it should be possible to realize it even in real material systems like transition metal dichalcogenide (TMD) heterostructres or graphene heterostructures where one can excite each layer with a infrared laser, which mimics the effect of time dependent potential on each layer which can be made out of phase by a time lag. It should also be possible to realize this drive in artificial gate controlled superlattices ~\cite{Cano}.
\begin{figure}
  \begin{center}
    \includegraphics[width=3.0in,angle=0]{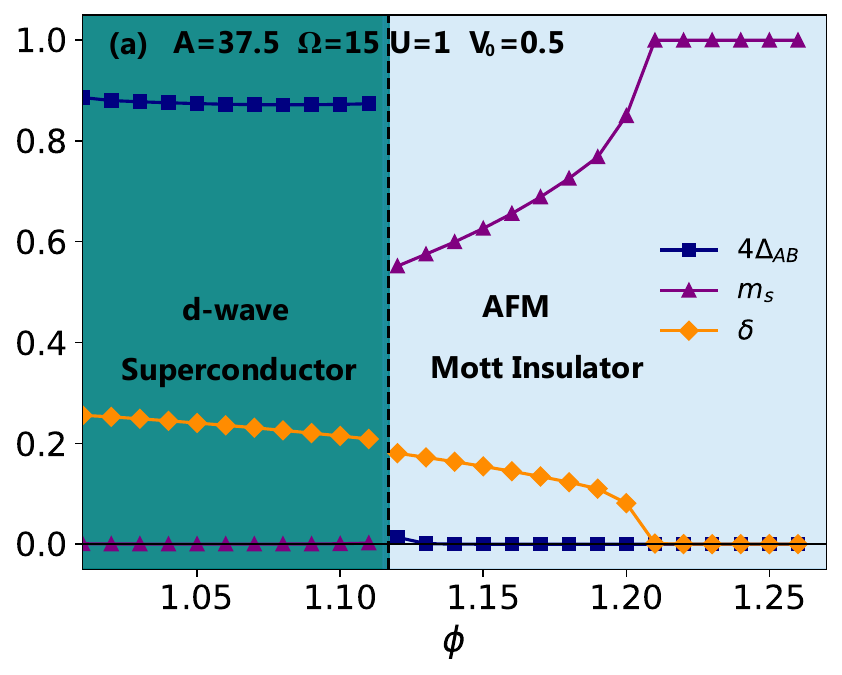}
  \includegraphics[width=3.25in,angle=0]{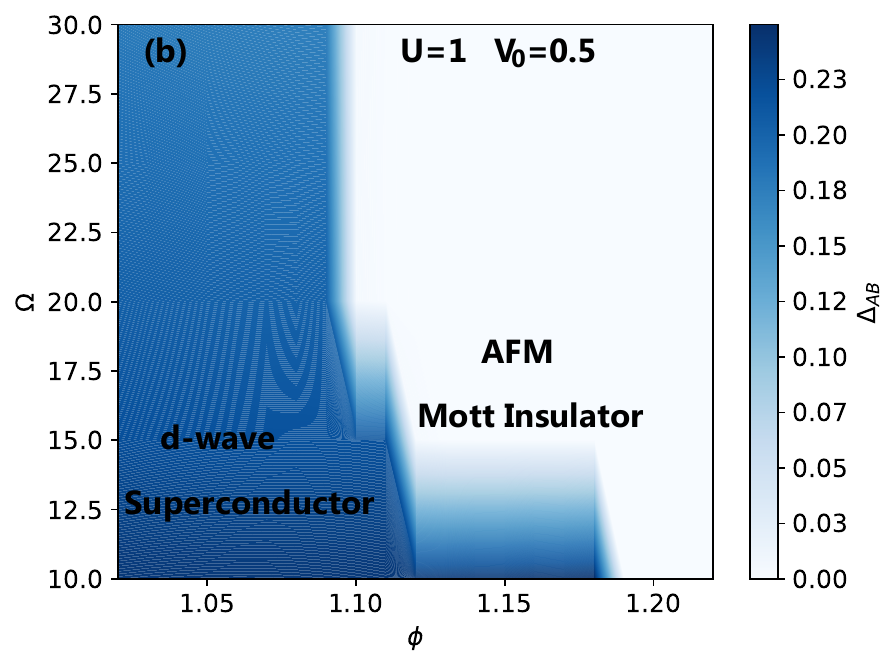}
    \includegraphics[width=5.5in,angle=0]{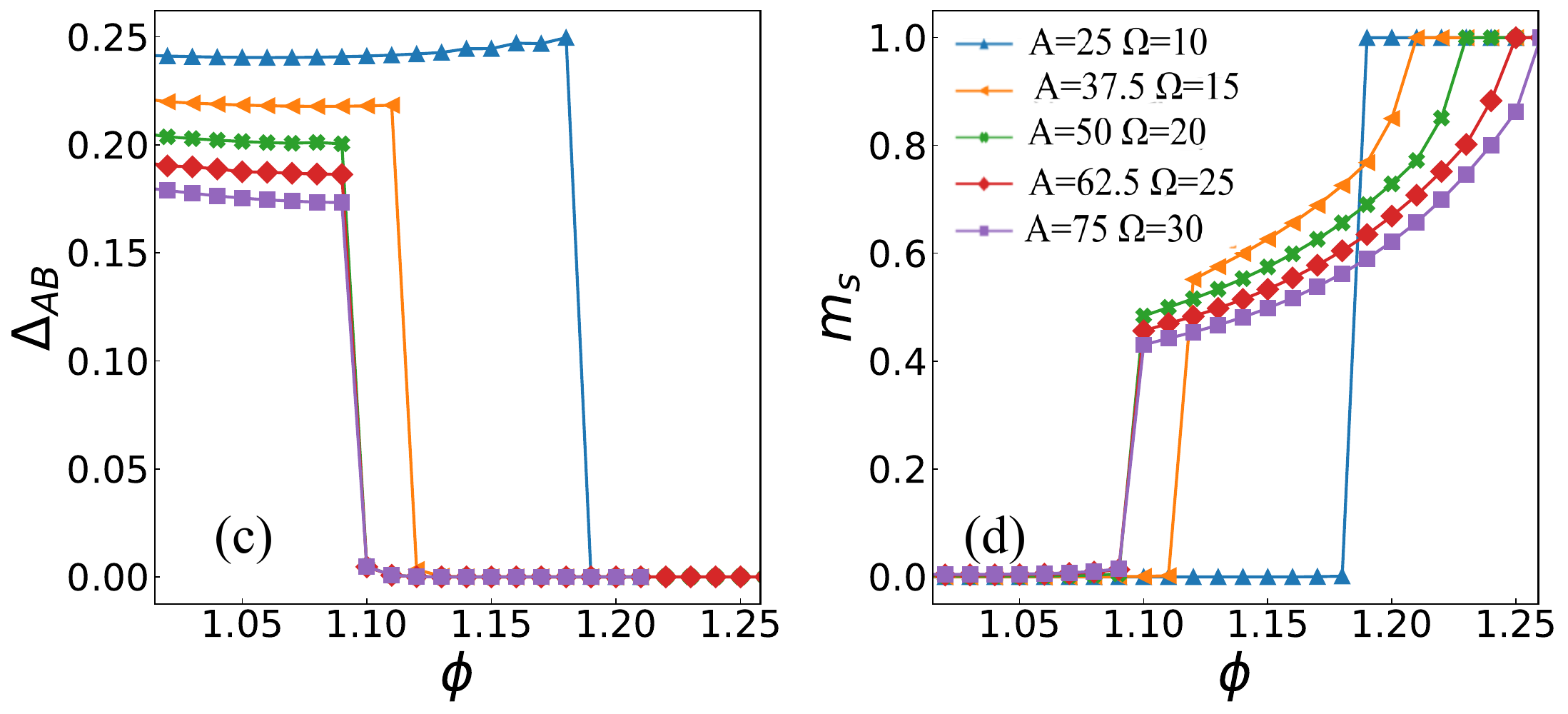}

          \caption{Top panels: Phase diagram of periodically driven Hubbard model with drive frequency $\Omega=15$ and drive amplitude $A=2.5\Omega$. As the phase $\Phi$ is tuned, at a threshold $\Phi_c$ where $U_{eff}+2V_{eff}\gg 1$, and the higher range hoppings $t^\prime_{eff}$ are significantly large, the system shows d-wave superconductivity. On increasing $\Phi$ further, dwave pairing amplitude goes to zero with a sharp drop and the staggered magnetization turns on such that the system transits into an AFM Mott Insulator. The phase diagram in the right panel shows the colormap of d-wave pairing amplitude in $\Omega-\Phi$ plane. The d-wave superconductivity exists for a broad range of drive frequencies(with $A=2.5\Omega$). Bottom panels shows the basic plots of the d-wave pairing amplitude $\Delta_{AB}$ and the staggered magnetization $m_s$ vs $\Phi$ for a range of drive parameters.          
          The data shown are for $U=1.0t_0$ and staggered potential $V_0=0.5t_0$ in the undriven Hamiltonian.}
          \label{phase_diag}
          \end{center}
          \end{figure}

\noindent
\textbf{The strong correlation limit of the model:} 
Close to the dynamical localization point, $U,V\gg t_{eff},t^\prime_{ind}$ resulting in an effectively strongly interacting system even though the bare $U$ and $V_0$ (if started with finite $V_0$) are comparable to $t_0$. This is depicted in Fig.~\ref{Fig2}, which presents various couplings in $H_F$ as functions of the phase $\Phi$ for a wide range of drive frequencies with drive amplitude $A=2.5\Omega$. The data shown is for initial $U=1.0t_0$ and $V_0=0.1t_0$. Even before the system hits dynamical localization point, nearest neighbour hopping $t_{eff}$ gets sufficiently suppressed for a broad range of $\Phi$, e.g., starting from $\Phi=0.9$ for the parameters shown in Fig~\ref{Fig2}. This results in enhanced values of $U_{eff}+2V_{eff}$ which is the energy cost of having a doublon on A sublattice. Similarly, the effective staggered potential $2V_{eff}=2(V_0+V_{ind})/t_{eff}$ which is the energy cost of hopping a holon from site A to B gets enhanced for a wide range of $\Phi$ as shown in Fig.~\ref{Fig2}. Bottom panels present the colorplots of the drive induced higher range hoppings $t^\prime_{eff}=t^\prime/t_{eff}$ and the correlated hopping $t^{eff}_{cor}=t_{cor}/t_{eff}$. Second neighbour hopping induced by the drive is really essential for a stable SC phase. Though this staggered hopping does not break particle-hole symmetry, but in the strong correlation regime, it can frustrate the AFM order via second neighbour spin-exchange interaction as shown below. 

We would like to emphasize that the second and third neighbour hoppings induced by the periodic drive, which are really crucial to stabilize the superconductivity, are not generally easy to be implemented in cold atom systems without the drive. In fact almost all cold atom experiments on Fermi Hubbard model or variants of it , have been done without the second neighbour hopping term. Periodically driving the Hubbard model naturally induces these interesting additional terms in the Hamiltonian which are crucial to obtain exotic phases like the unconventional superconductivity. 
In a half-filled system, due to high energy cost of the doublons on A sublattice, doublons should be projected out from the low energy Hilbert space of A sublattice and holes from the low energy Hilbert space of B sublattice.
The effective low energy Hamiltonian in this limit is obtained by transforming the Hamiltonian in Eqn.[\ref{HF1}] via Schriefer-Wolf transformation which eliminates the processes that interconnect the low energy and high energy sectors.   
\begin{figure}
  \begin{center}
\includegraphics[width=6.0in,height=1.9in]{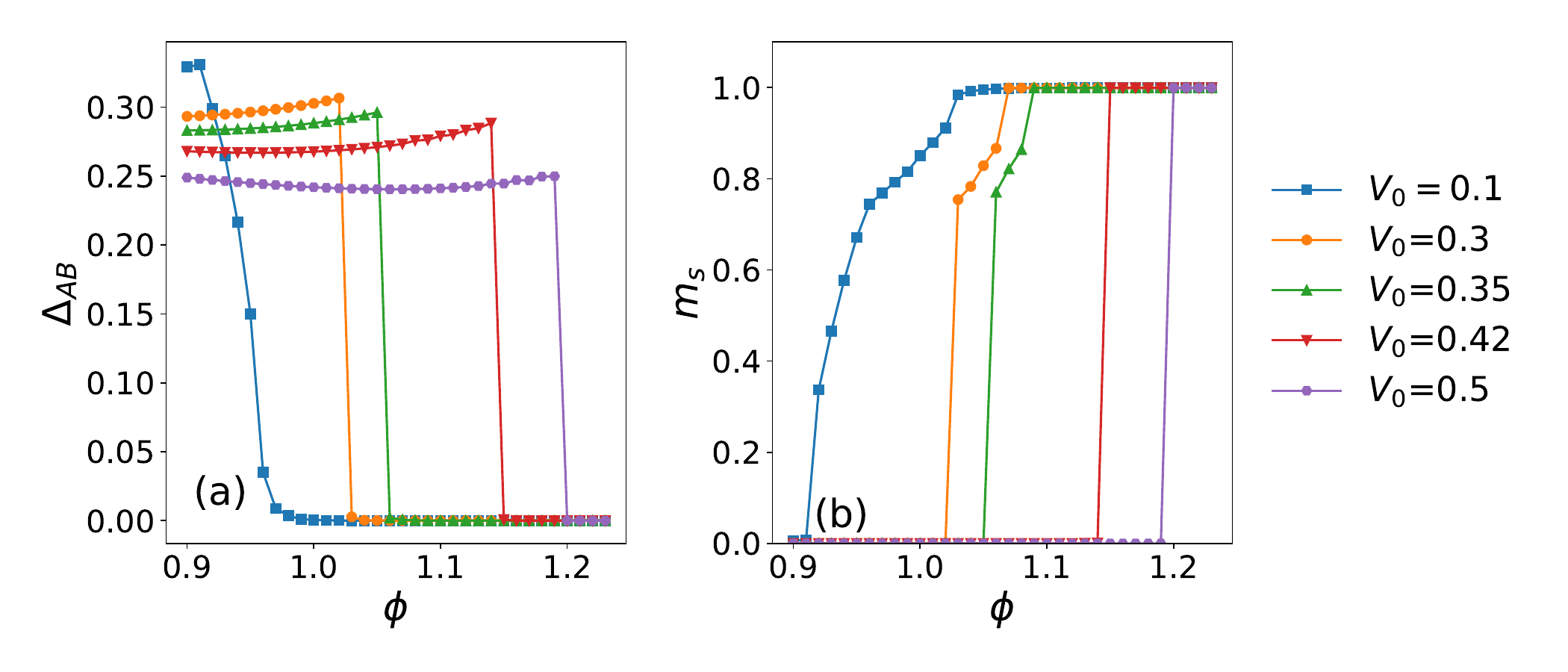}
   \includegraphics[width=3.5in,angle=0]{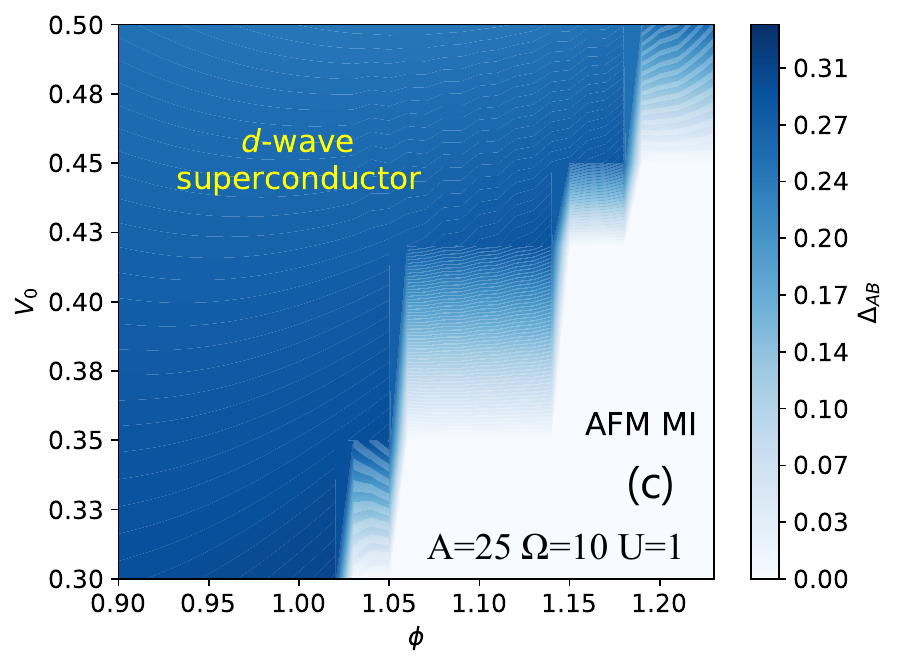}
  \includegraphics[width=5.5in,angle=0]{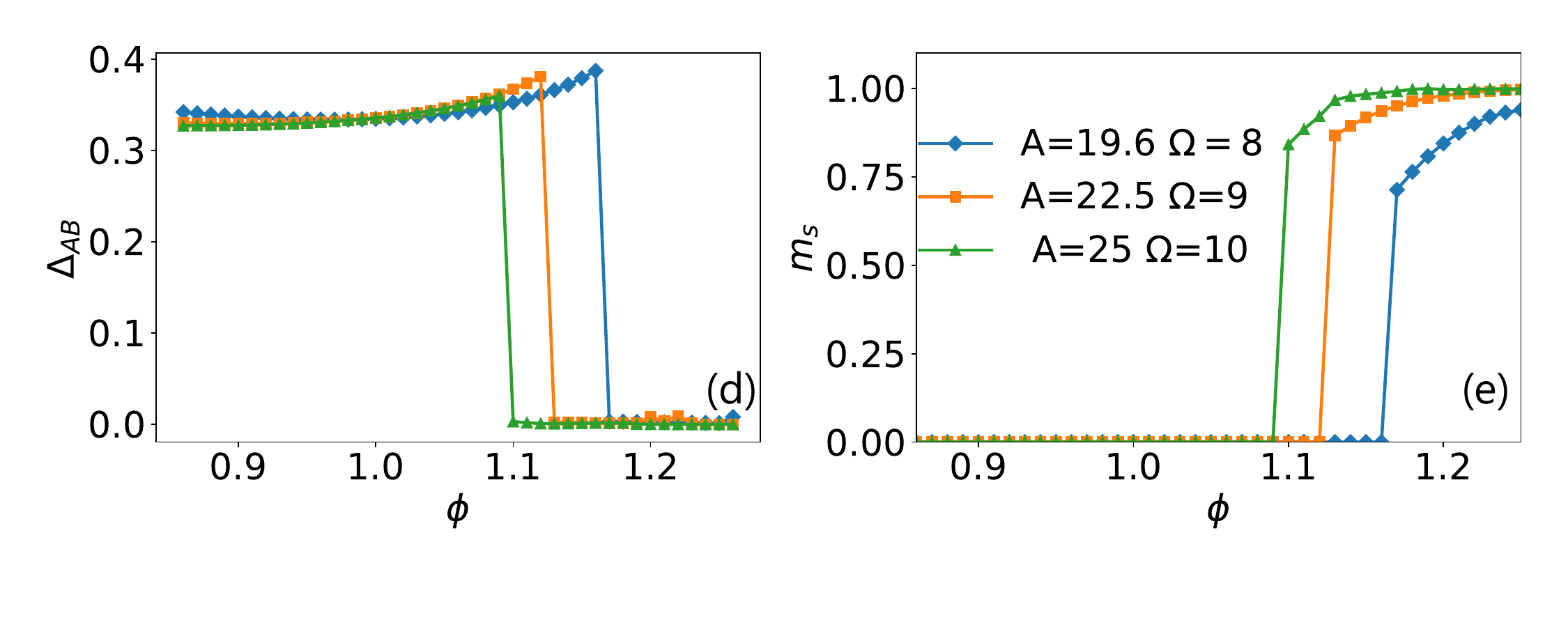}

          \caption{Panel[a] shows the d-wave pairing amplitude $\Delta_{AB}$ vs $\Phi$ for a range of initial values of the staggered potential $V_0$ for a fixed value of the Hubbard interaction $U=1.0$. Panel [b] shows the staggered magnetization $m_s$ vs $\Phi$ for various values of $V_0$ for $U=1.0$. Panel[c] shows the consolidated phase diagram in $V_0-\Phi$ plane for the periodic drive with $\Omega=10$ and $A=2.5\Omega$ based on the results shown in panel[a]. The d-wave superconducting phase survives for a broad range of staggered potential $V_0$. Panel[c] and [d] shows the results for the system with $V_0=0$ and $U=0.55t_0$. The d-wave superconductivity survives even in this case for a range of drive parameters.} 
          \label{diff_V0}
          \end{center}
          \end{figure} 
The effective Hamiltonian in the projected Hilbert space is given as below:
\begin{widetext}
\begin{equation}
\begin{split}
\mathcal{H}_{eff}/t_{eff}=\mathcal{P}\bigg[-\sum_{\langle ij \rangle ,\sigma}c_{iA\sigma}^{\dagger}c_{jB\sigma}+h.c.+2t^\prime_{eff}\sum_{(ij)}c_{iA\sigma}^{\dagger}c_{jA\sigma}-c_{iB\sigma}^{\dagger}c_{jB\sigma}\bigg]\mathcal{P} \\+\mathcal{P}\bigg[t^\prime_{eff}\sum_{((ij))}c_{iA\sigma}^{\dagger}c_{jA\sigma}-c_{iB\sigma}^{\dagger}c_{jB\sigma}
+J\sum_{\langle ij \rangle} \big[S_{iA}.S_{jB}-\frac{n_A(2-n_B)}{4}\big]\bigg]\mathcal{P}\\+
\mathcal{P}\bigg[4J^\prime \sum_{(ij)} \big[S_{iA}.S_{jA}-\frac{n_{iA}n_{jA}}{4}\big]+4J^\prime\sum_{(ij)} \big[S_{iB}.S_{jB}-\frac{(2-n_{iB})(2-n_{jB})}{4}\big]\bigg]\mathcal{P}\\
+\mathcal{P}\bigg[J^\prime \sum_{((ij))} \big[S_{iA}.S_{jA}-\frac{n_{iA}n_{jA}}{4}\big]+J^\prime\sum_{((ij))} \big[S_{iB}.S_{jB}-\frac{(2-n_{iB})(2-n_{jB})}{4}\big]\bigg]\mathcal{P}\\
+ t_{\mathrm{cor}}^{eff} \sum_{\sigma,<ij>}\mathcal{P}
\big[(n_{iA\bar{\sigma}} - n_{j B \bar{\sigma}})
c^{\dagger}_{iA\sigma}c_{j B\sigma}\big] \mathcal{P} +H_0+H_d-\mu\sum_{i}n_{i}
\end{split}
\label{tJ}
\end{equation}
\end{widetext}

Here,$J=2/(U_{eff}+2V_{eff})$ and $J^\prime=4(t^\prime_{eff})^2/U_{eff}$ are inter-sublattice and intra-sublattice spin-exchange interactions. $\mathcal{P}$ is the projection operator that projects out doublons from sublattice A and holes from sublattice B. $H_0= \frac{U_{eff}-2V_{eff}}{2}\sum_{i}\big[n_{iB\uparrow}n_{iB\downarrow}+(1-n_{iA\uparrow})(1-n_{iA\downarrow})\big]$ and the dimer term $H_d= -\frac{1}{2V_{eff}}\sum_{\langle ij \rangle ,\sigma}\mathcal{P}\big[(1-n_{jB\bar{\sigma}})(1-n_{iA})+(n_{jB}-1)n_{iA\sigma}\big]\mathcal{P}$. The projection constraints in the effective low energy Hamiltonian are handled within the Gutzwiller approximation~\cite{Ogata} by renormalizing various coefficients in $\mathcal{H}_{eff}$ by appropriate Gutzwiller factors (see details in SM). Gutzwiller approximations~\cite{edegger-advphysics,Ogata,Anwesha1,Anwesha3} of the sort we use  have been well vetted against quantum Monte Carlo calculations~\cite{Arun,zhang,Watanabe} and dynamical mean field theory~\cite{Anwesha1}. Details of this Gutzwiller approximation for the various terms in $H_{eff}$ are given in the Methods section. 

\noindent
{\bf{Methods:}} $H_{eff}$ acts on a projected Hilbert space which consists of states $|\Psi \rangle  = P| \Psi_{0} \rangle $  where the projection operator $P$ eliminates configurations with doublons from sublattice A and holons from sublattice B from unprojected state $|\Psi_0\rangle$.  We use here the Gutzwiller approximation to handle the projection constraints which maps various correlation functions in non-canonical fermionic operators in the projected Hilbert space to canonical operators in the unprojected Hilbert space weighted by appropriate Jastro factors. Thus, we obtain the kinetic energy $\langle c^\dagger_{i\alpha,\sigma}c_{j\beta,\sigma}\rangle\approx g_t(i\alpha,j\beta)\langle c^\dagger_{i\alpha,\sigma}c_{j\beta,\sigma}\rangle_0$. Here $\alpha,\beta=A,B$ are the labels for two sublattices. The spin correlation $\langle S_{i}· S_{j}\rangle \approx g_s(i,j)\langle S_i· S_j\rangle_0$. The local Gutzwiller factors given are by $g_{tA\sigma}=\frac{2\delta}{(1+\delta-\sigma m_A)}$ for intra-sublattice hopping on A sublattice,$g_{tB\sigma}=\frac{2\delta}{(1+\delta+\sigma m_B)}$ for intra-sublattice hopping on B sublattice, and for nearest neighbour intersublattice hopping $g_t=\sqrt{g_{A\sigma} g_{B\sigma}}$. The Gutzwiller factor for spin-exchange coupling is $g_{s\alpha \beta}=\frac{4}{\sqrt((1+\delta)^2-m_{\alpha}^2)((1+\delta)^2-m_{\beta}^2))}$. We then obtain the Gutzwiller renormalized hamiltonian as the one in which all operators are replaced by the corresponding operators in unprojected space but are multiplied by above mentioned Jastro factors based on Gutzwiller approximation. 


We solve the low energy hamiltonian by renormalized Bogoliubov-deGennes approach. In which the spin-exchange terms and dimer terms are decomposed into particle-particle and particle-hole channels and calculate the expectation values of the mean fields (a) pairing amplitude $\Delta_{AB}=\langle c_{A\uparrow}^{\dagger}c_{B\downarrow}^{\dagger} \rangle-\langle c_{A\downarrow}^{\dagger}c_{B\uparrow}^{\dagger} \rangle$
(b) inter sublattice fock shift $\chi_{AB\sigma}=\langle c_{A \sigma}^{\dagger}c_{B \sigma} \rangle$ 
(c) intra sublattice fock shifts on A(B) sublattice $\chi_{\alpha \alpha \sigma}=\langle c_{i\alpha\sigma}^{\dagger}c_{i+2x/2y \alpha \sigma} \rangle$, and $\chi_{\alpha \alpha \sigma}=\langle c_{i\alpha\sigma}^{\dagger}c_{i \pm x \pm y \alpha \sigma} \rangle$
We study the Spin asymmetric calculation. in which the magnetic order parameters like staggered magnetization $m_f=(m_A+m_B)/2$ and staggered magnetization $m_s=(m_B-m_A)/2$ where $m_{s \alpha}=n_{\alpha \uparrow}-n_{\alpha \downarrow}$ are allowed to have finite expectation values. 

\noindent
{\bf{Results}}:
We solve The Hamiltonian in Eq.[~\ref{HF1}] using spin-asymmetric Gutzwiller renormalized Bogoliubov-Degennes (BdG) approach to explore the possibility of unconventional superonconductivity in this periodically driven system. We investigated the possibility of inter-sublattice $d_{x^2-y^2}$-wave and extended s-wave pairing  and  intra-sublattice pairing of $d_{xy}$ on 2nd neighbours and $d_{x^2-y^2}$ on 3rd neighbours though only $d_{x^2-y^2}$ inter-sublattice pairing is found to be stable. We only looked for the superconducting phases in the regime where $U_{eff}+2V_{eff}\gg 1$ where our approach to the Floquet Hamiltonian is eligible. Details of renormalized BdG theory are provided in the SM. 

Fig.~\ref{phase_diag} shows the most important findings of our work. Panel[a] in Fig.~\ref{phase_diag} shows the phase diagram for a fixed drive frequency and amplitude, both being in the off-resonant regime. d-wave superconductivity is characterized by the finite value of the d-wave pairing amplitude $\Delta_{AB} =\sum_k\gamma(k)\langle c^\dagger_{kA\uparrow}c^\dagger_{-kB\downarrow}\rangle$ with $\gamma(k)=\cos(k_x)-\cos(k_y)$. As shown in Panel [a], the d-wave superconductivity survives for a wide range of the phase $\Phi$ of the drive  where $t^\prime_{eff}$ is large enough to generate sufficient strength of intra-sublattice spin-exchange interactions which frustrates the antiferromagnetic order induced by nearest neighbour spin-exchange interactions. At $\Phi_c$, $\Delta_{AB}$ drops to zero and the AFM order sets in resulting in a finite value of the staggered magnetization $m_s = m_B-m_A$. This is the AFM Mott insulator phase with a gap in the single particle excitation spectrum as shown in the SM. As $\Phi$ is tuned further beyond $\Phi_c$, $m_s$ increases and eventually saturates to one in the extremely correlated regime of the Floquet Hamiltonian. This is also the regime where $t^\prime_{eff}$ has almost decayed to zero. Throughout the superconducting phase and also a part of the AFM Mott insulator phase, there is a finite density difference $\delta=n_B-n_A$ between the two sublattices. 

Superconducting phase is very robust against the drive parameters and survives for a wide range of drive frequency and amplitudes as shown in the panel[b] of Fig.~\ref{phase_diag} which shows a colormap of $\Delta_{AB}$ in $\Omega-\Phi$ plane for $A=2.5\Omega$. As the drive frequency is lowered, there is an increase in the range of $\Phi$ values for which the system shows superconductivity which is because of the enhanced value of $t^{\prime}_{eff}$ for lower drive frequencies as shown in Fig.~\ref{Fig2}. Bottom panels in Fig.\ref{phase_diag} presents the basic data based on which the phase diagram in panel [a] and [b] has been drawn. As the drive frequency decreases, though still being much larger than intrinsic energy scale of the undriven Hamiltonian, the range of superconducting phase in $\Phi$ increases. Staggered magnetization turns on at a larger value of $\Phi$ now and $m_s$ at $\Phi_c$ is larger for the system with higher drive frequency.

So far we have presented the phase diagram for a fixed value of initial parameters in the Hamiltonian, namely, $U=1.0t_0$ and $V_0=0.5t_0$. We further explore the dependence of the superconducting phase on the initial value of the staggered potential $V_0$ and demonstrate that the SC can be stabilized even for $V_0=0$ if the drive parameters are in appropriate regime.
Fig.~[\ref{diff_V0}], shows the results for various values of $V_0$ for $U_0=1.0t_0$. The d-wave superconductivity is realized not only for $V_0 = U/2$ but for a wide range of $V_0$ values. For a given set of drive parameters, the range in phase $\Phi$ for which the system shows superconductivity decreases monotonically as $V_0$ decreases. $V_0$
 has two interesting effects on $H_F$. First of all the drive induced contribution to nearest neighbour hopping $t_{ind}$ linearly depends on $V_0$ which eventually results in smaller $t_{eff}$ for smaller values of staggered potential $V_0$. Secondly, as $t_{eff}$ decreases with decrease in $V_0$, this results in effective enhancement of $t^\prime_{eff}$ which supports superconductivity. For large values of $|U_{eff}-2V_{eff}|$, the system will loose its delicate charge dynamics arising from the density difference and the difference in projection constraints on the two sublattices, which goes against superconducting phase.   The combined results of all these effect is that the width of the superconducting phase in $\Phi$ decreases as $V_0$ decreases. The analysis of various initial values of staggered potential  are shown in the phase diagram of panel [c] and details of couplings and various hoppings in $H_F$ are provided in the SM. 

Although for a given set of periodic drive, the width of the SC phase decreases as $V_0$ decreases, it is still possible to realize superconductivity even in the system with no initial staggered potential $V_0=0$, with appropriate choice of drive parameters. Now the only contribution to  $V_{eff}$ is from the staggered potential induced by the drive $V_{ind}$. Even in this case for a range of drive parameters, $U_{eff}+2V_{eff}$ can be sufficiently large compared to the hopping terms resulting in the onset of d-wave superconductivity as shown in panels [c] of the Fig.~[\ref{diff_V0}] for $U=0.55t_0$. But this superconducting phase is restricted to lower values of the drive frequency, still being in the off-resonant regime. As the drive frequency is increased further, the drive induced staggered potential decreases. As discussed before, for the system to have density difference between the two sublattices and have different projection constraints on the two sublattices, it is essential that $V_{eff} \sim U_{eff}$. Because only in this regime, even at half-filling, the system still has charge dynamics which is essential for realizing conducting phases like d-wave superconductor.  

          \begin{figure}
  \begin{center}
    \includegraphics[width=4.9in,angle=0]{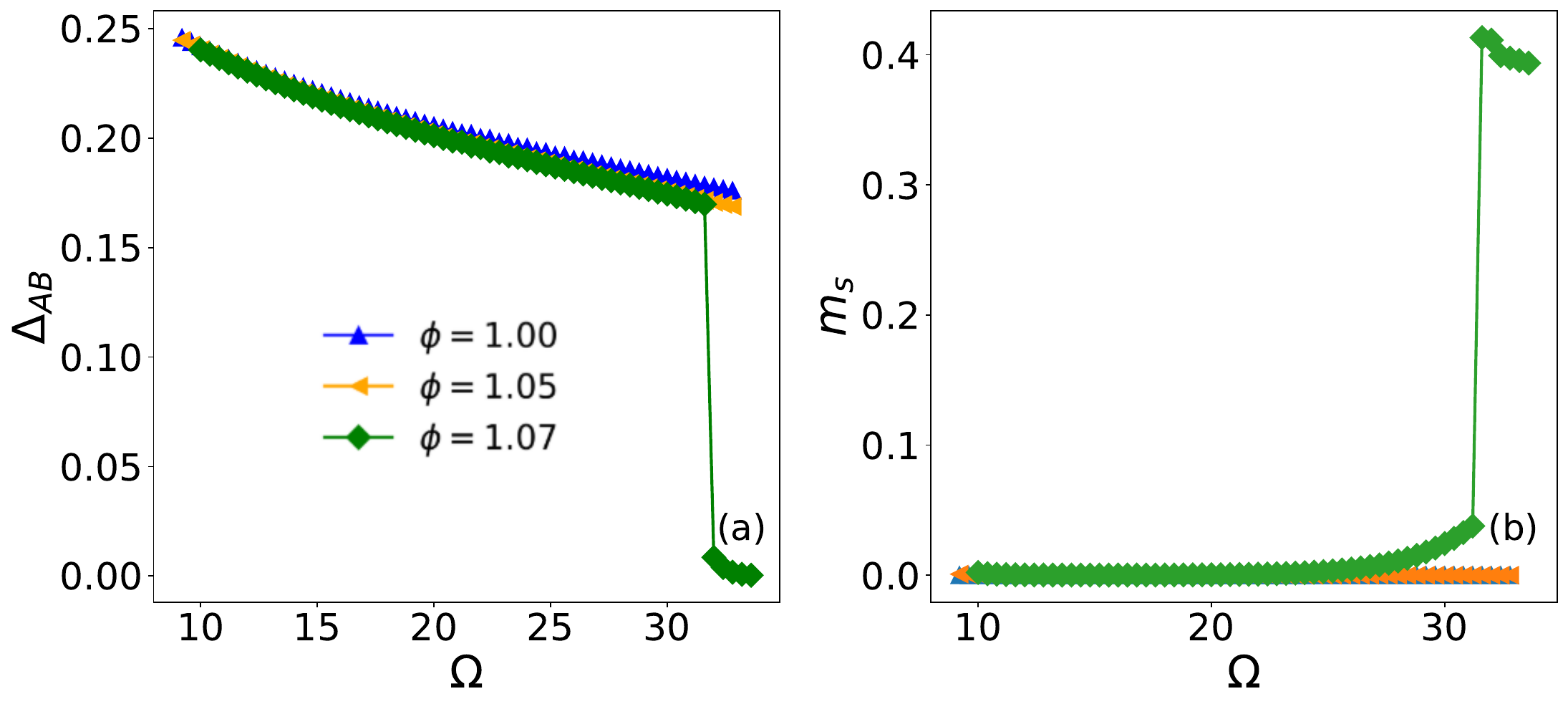}
          \caption{Left Panel shows d-wave pairing amplitude vs the drive frequency $\Omega$ for various values of the phase $\Phi$. The right panel shows the staggered magnetization $m_s$ vs phase $\Omega$ of the drive for the same set of $\Phi$ values. The data shown are for $A= 2.5\Omega$, $U=1.0t_0$ and $V_0=0.5$.}
          \label{fixPhi}
          \end{center}
          \end{figure}

\noindent
{{\bf Alternate Driving Protocol:}}
We have also studied the system in an alternate drive protocol in which the phase $\Phi$ is kept fixed while the amplitude and frequency of the drive are tuned keeping the ratio $A/\Omega$ fixed. Though the contribution of the Bessel function $\mathcal{J}_0(A/\Omega\sin(\Phi))$ in the effective nearest neighbour hopping is fixed for a given value of $\Phi$ and $A/\Omega$, the drive induced nearest neighbour hopping $t_{ind}$ and higher range hoppings $t^\prime_{ind}$ decrease as $\Omega$ increases. Therefore, in this protocol, superconductivity survives for a broad range of the drive frequencies (shown in Fig.~[\ref{fixPhi}]) as long as $J^\prime$ is large enough to frustrate the AFM order and stabilize the SC. Details of various drive induced hoppings and effective couplings in $H_F$ are provided in the SM. 

The method used to investigate superconductivity in this work has been vetted against various methods like slave boson approach and variational quantum Monte Carlo methods~\cite{edegger-advphysics, Watanabe} and to a large extent with dynamical mean field theory~\cite{Anwesha2}. Still it would be crucial to explore the system under study with more accurate methods like tensor-networks and density matrix renormalization group~\cite{TN1,TN2}. 

\noindent
{\bf{Conclusions:}} In summary, we have established a non-equilibrium framework for realizing unconventional superconductivity at commensurate filling, bypassing the long-standing challenges associated with chemical doping and quenched disorder. By leveraging the high-frequency prethermal stability of periodically driven many-body systems, we demonstrated that a bipartite Hubbard model and its variants can be dynamically tuned into a d-wave superconducting state. The mechanism hinges on a drive-induced renormalization of kinetic scales and the emergence of a staggered potential, which creates a "local doping" effect that sustains carrier dynamics despite a global half-filled density.
The dynamical stabilization of these phases is particularly promising for a broad spectrum of quantum simulators. Beyond the ultracold atom platforms discussed here~\cite{Esslinger1,Esslinger2,Bloch_Fermi_Hubbard}, the proposed Floquet mechanism provides a roadmap for engineering correlations in diverse architectures like quantum dot arrays~\cite{Quantum_Dot_array_Fermi_Hubbard}, artificial superlattices~\cite{Cano} and Google Sycamore chip~\cite{Google_Fermi_Hubbard}. It should also be possible to realize the periodic drive induced superconductivity in real material systems like transition metal-dichalogenide (TMD) heterostructures or graphene heterostructures as discussed before.

The implications of this work extend significantly into the realm of quantum technologies. Unconventional superconducting phases are essential components of superconducting qubit arrays and next-generation quantum computers~\cite{qubits} Currently, the performance of these architectures is heavily constrained by randomness in Josephson energies—a direct result of material inhomogeneity~\cite{Berke_qubit,Gao_qubit}. By stabilizing superconductivity at half-filling without the introduction of chemical disorder, our protocol offers a pathway toward more homogeneous, reproducible superconducting platforms.
Ultimately, this Floquet-mediated route underscores the potential of non-equilibrium protocols to unlock hidden phases in commensurate parent compounds. By providing a "disorder-free" alternative for on-demand quantum material design, our work establishes a robust foundation for optimizing the coherence and scalability of future quantum computing and simulation technologies.
\\
\\
\noindent
{\bf{Acknowledgments}}
\\
 We would like to acknowledge H.R. Krishnamurthy and Arnab Das for various stimulaitng discussions on periodically driven systems. AG would like to acknowledge Anwesha Chattopadhyay for discussions and past collaboration on a related project.

\bibliography{HM_drive.bib}

@article{light_sc_cuprates,
doi={10.1126/science.1197294},
author = {D. Fausti  and R. I. Tobey  and N. Dean  and S. Kaiser  and A. Dienst  and M. C. Hoffmann  and S. Pyon  and T. Takayama  and H. Takagi  and A. Cavalleri },
title = {Light-Induced Superconductivity in a Stripe-Ordered Cuprate},
journal = {Science},
volume = {331},
number = {6014},
pages = {189-191},
year = {2011},
URL = {https://www.science.org/doi/abs/10.1126/science.1197294},
eprint = {https://www.science.org/doi/pdf/10.1126/science.1197294},
}

@article{light_cuprates2,
  title = {Emergence of light-induced superconducting-like state from the charge density wave state in high-${T}_{c}$ cuprate superconductors},
  author = {Nishida, Morihiko and Song, Dongjoon and Hallas, Alannah M. and Eisaki, Hiroshi and Shimano, Ryo},
  journal = {Phys. Rev. B},
  volume = {110},
  issue = {22},
  pages = {224515},
  numpages = {12},
  year = {2024},
  month = {Dec},
  publisher = {American Physical Society},
  doi = {10.1103/PhysRevB.110.224515},
  url = {https://link.aps.org/doi/10.1103/PhysRevB.110.224515}
}

@article{Cavalleri_PRX,
  title = {Pump Frequency Resonances for Light-Induced Incipient Superconductivity in ${\mathrm{YBa}}_{2}{\mathrm{Cu}}_{3}{\mathrm{O}}_{6.5}$},
  author = {Liu, B. and F\"orst, M. and Fechner, M. and Nicoletti, D. and Porras, J. and Loew, T. and Keimer, B. and Cavalleri, A.},
  journal = {Phys. Rev. X},
  volume = {10},
  issue = {1},
  pages = {011053},
  numpages = {8},
  year = {2020},
  month = {Mar},
  publisher = {American Physical Society},
  doi = {10.1103/PhysRevX.10.011053},
  url = {https://link.aps.org/doi/10.1103/PhysRevX.10.011053}
}

@article{fulleren_sc,
title={Evidence for metastable photo-induced superconductivity in K3C60},
author={Budden, M. and Gebert, T. and Buzzi et al, M.},
Journal={Nature Physics},
volume={17},
page={611},
year={2021},
doi={https://doi.org/10.1038/s41567-020-01148-1},
url={https://www.nature.com/articles/s41567-020-01148-1#citeas}
}

@article{fulleren2,
author={Cantaluppi, A. and Buzzi, M. and Jotzu et al, G.},
title={Pressure tuning of light-induced superconductivity in $K_3C_{60}$.},
journal={Nature Phys.},
volume={14}, 
page={837–841},
year={2018},
doi={https://doi.org/10.1038/s41567-018-0134-8},
}

@article{fullerene1,
author={Mitrano, M. and Cantaluppi, A. and Nicoletti et al, D.},
title={Possible light-induced superconductivity in $K_3C_{60}$ at high temperature.},
journal={Nature},
volume={530}, 
page={461–464},
year={2016},
doi={https://doi.org/10.1038/nature16522},
}

@article{Coulthard,
  title = {Enhancement of superexchange pairing in the periodically driven Hubbard model},
  author = {Coulthard, J. R. and Clark, S. R. and Al-Assam, S. and Cavalleri, A. and Jaksch, D.},
  journal = {Phys. Rev. B},
  volume = {96},
  issue = {8},
  pages = {085104},
  numpages = {13},
  year = {2017},
  month = {Aug},
  publisher = {American Physical Society},
  doi = {10.1103/PhysRevB.96.085104},
  url = {https://link.aps.org/doi/10.1103/PhysRevB.96.085104}
}

@article{Chiral_SC,
  title = {Flexible Control of Chiral Superconductivity in Optically Driven Nodal Point Superconductors with Antiferromagnetism},
  author = {Ning, Zhen and Ma, Da-Shuai and Zeng, Junjie and Xu, Dong-Hui and Wang, Rui},
  journal = {Phys. Rev. Lett.},
  volume = {133},
  issue = {24},
  pages = {246606},
  numpages = {7},
  year = {2024},
  month = {Dec},
  publisher = {American Physical Society},
  doi = {10.1103/PhysRevLett.133.246606},
  url = {https://link.aps.org/doi/10.1103/PhysRevLett.133.246606}
}

@article{zeng_sc,
  title = {Inducing and controlling superconductivity in the Hubbard honeycomb model using an electromagnetic drive},
  author = {Kumar, Umesh and Lin, Shi-Zeng},
  journal = {Phys. Rev. B},
  volume = {103},
  issue = {6},
  pages = {064508},
  numpages = {10},
  year = {2021},
  month = {Feb},
  publisher = {American Physical Society},
  doi = {10.1103/PhysRevB.103.064508},
  url = {https://link.aps.org/doi/10.1103/PhysRevB.103.064508}
}

@article{TGA,
author={Anan, T. and Morimoto, T.  and Kitamura, S. },
title={Time-dependent Gutzwiller simulation of Floquet topological superconductivity.},
Journal={Commun Phys.},
volume={7},
page={99},
year={2024},
url={https://doi.org/10.1038/s42005-024-01586-w},
}

@article{qubits,
   author = "Kjaergaard, Morten and Schwartz, Mollie E. and Braumüller, Jochen and Krantz, Philip and Wang, Joel I.-J. and Gustavsson, Simon and Oliver, William D.",
   title = "Superconducting Qubits: Current State of Play", 
   journal= "Annual Review of Condensed Matter Physics",
   year = "2020",
   volume = "11",
   number = "Volume 11, 2020",
   pages = "369-395",
   doi = "https://doi.org/10.1146/annurev-conmatphys-031119-050605",
   url = "https://www.annualreviews.org/content/journals/10.1146/annurev-conmatphys-031119-050605",
   publisher = "Annual Reviews",
   issn = "1947-5462",
   type = "Journal Article",
   }

@article{Berke_qubit,
author={Berke, C. and Varvelis, E. and Trebst et al, S.},
title={Transmon platform for quantum computing challenged by chaotic fluctuations.},
journal={Nat Commun},
volume={13}, 
page={2495},
year={2022},
url={https://doi.org/10.1038/s41467-022-29940-y},
}

@article{Gao_qubit,
   author={Gao, R. and Wu, F. and Sun et al, H.},
   title={The effects of disorder in superconducting materials on qubit coherence.},
   journal={Nat Commun},
   volume={16},
   page={3620},
   year={2025},
   url={https://doi.org/10.1038/s41467-025-58745-y},
   }

@article{graphene_AQH,
author={McIver, J.W. and Schulte, B. and Stein et al, F.},
title={Light-induced anomalous Hall effect in graphene},
Journal={Nature Physics},
volume={16},
page={38},
year={2020},
doi={https://doi.org/10.1038/s41567-019-0698-y},
url={https://www.nature.com/articles/s41567-019-0698-y}
}

@article{AQHE,
author={Xu, H. and Zhou, J. and Li, J.},
title={Light-Induced Quantum Anomalous Hall Effect on the 2D Surfaces of 3D Topological Insulators},
journal={Adv. Sciences},
volume={8},
pages = {2101508},
year={2021},
doi = {https://doi.org/10.1002/advs.202101508},
url = {https://advanced.onlinelibrary.wiley.com/doi/abs/10.1002/advs.202101508},
}

@article{MBL_light,
  title = {Activating Many-Body Localization in Solids by Driving with Light},
  author = {Lenar\ifmmode \check{c}\else \v{c}\fi{}i\ifmmode \check{c}\else \v{c}\fi{}, Zala and Altman, Ehud and Rosch, Achim},
  journal = {Phys. Rev. Lett.},
  volume = {121},
  issue = {26},
  pages = {267603},
  numpages = {6},
  year = {2018},
  month = {Dec},
  publisher = {American Physical Society},
  doi = {10.1103/PhysRevLett.121.267603},
  url = {https://link.aps.org/doi/10.1103/PhysRevLett.121.267603},
}

@article{AFMorder,
  title = {Ultrafast light-induced long-range antiferromagnetic correlations in paramagnets},
  author = {Amato, Lorenzo and M\"uller, Markus},
  journal = {Phys. Rev. B},
  volume = {109},
  issue = {5},
  pages = {054424},
  numpages = {11},
  year = {2024},
  month = {Feb},
  publisher = {American Physical Society},
  doi = {10.1103/PhysRevB.109.054424},
  url = {https://link.aps.org/doi/10.1103/PhysRevB.109.054424}
}

@article{charge_order,
author={Kogar, A. and Zong, A. and Dolgirev et al, P.E.},
title={Light-induced charge density wave in LaTe3},
volume={16},
journal={Nature Physics},
pages={159},
year={2020},
doi={https://doi.org/10.1038/s41567-019-0705-3},
url={https://www.nature.com/articles/s41567-019-0705-3},
}

@article{LDM_PRE,
  title = {Equilibrium states of generic quantum systems subject to periodic driving},
  author = {Lazarides, Achilleas and Das, Arnab and Moessner, Roderich},
  journal = {Phys. Rev. E},
  volume = {90},
  issue = {1},
  pages = {012110},
  numpages = {6},
  year = {2014},
  month = {Jul},
  publisher = {American Physical Society},
  doi = {10.1103/PhysRevE.90.012110},
  url = {https://link.aps.org/doi/10.1103/PhysRevE.90.012110}
}

@article{Alessio2014,
  title = {Long-time Behavior of Isolated Periodically Driven Interacting Lattice Systems},
  author = {D'Alessio, Luca and Rigol, Marcos},
  journal = {Phys. Rev. X},
  volume = {4},
  issue = {4},
  pages = {041048},
  numpages = {12},
  year = {2014},
  month = {Dec},
  publisher = {American Physical Society},
  doi = {10.1103/PhysRevX.4.041048},
  url = {https://link.aps.org/doi/10.1103/PhysRevX.4.041048}
}

@article{Kuwahara_Mori_Saito,
  title = {Floquet-Magnus theory and generic transient dynamics in periodically driven many-body quantum systems},
  author = {Kuwahara, Tomotaka and Mori, Takashi and Saito, Keiji},
  journal = {Annals of Physics},
  volume = {367},
  issue = {},
  pages = {96},
  numpages = {},
  year = {2016},
  month = {},
  publisher = {},
  doi = {https://doi.org/10.1016/j.aop.2016.01.012},
  url = {}
}

@article{Dima_Floquet_Prethermalization,
  title = {A Rigorous Theory of Many-Body Prethermalization for Periodically Driven and Closed Quantum Systems},
  author = {Abanin, Dimitry and De Roeck, Wojciech and Ho, Wen Wei and Huveneer, Francois},
  journal = {Commun. Math. Phys.},
  volume = {354},
  issue = {},
  pages = {809},
  numpages = {},
  year = {2017},
  month = {},
  publisher = {},
  doi = {10.1007/s00220-017-2930-x},
  url = {}
}

@article{Mori_Kuwahara_Saito_Prethermal_PRL,
  title = {Rigorous Bound on Energy Absorption and Generic Relaxation in Periodically Driven Quantum Systems},
  author = {Mori, Takashi and Kuwahara, Tomotaka and Saito, Keiji},
  journal = {Phys. Rev. Lett.},
  volume = {116},
  issue = {12},
  pages = {120401},
  numpages = {5},
  year = {2016},
  month = {Mar},
  publisher = {American Physical Society},
  doi = {10.1103/PhysRevLett.116.120401},
  url = {https://link.aps.org/doi/10.1103/PhysRevLett.116.120401}
}

@article{Ho_Mori_Abanin_Pretherm_Rev,
   title={Quantum and classical Floquet prethermalization},
   volume={454},
   ISSN={0003-4916},
   url={http://dx.doi.org/10.1016/j.aop.2023.169297},
   DOI={10.1016/j.aop.2023.169297},
   journal={Annals of Physics},
   publisher={Elsevier BV},
   author={Ho, Wen Wei and Mori, Takashi and Abanin, Dmitry A. and Dalla Torre, Emanuele G.},
   year={2023},
   month=jul, pages={169297} }

@article{Dima_Pretherm_PRL,
  title = {Exponentially Slow Heating in Periodically Driven Many-Body Systems},
  author = {Abanin, Dmitry A. and De Roeck, Wojciech and Huveneers, Fran\ifmmode \mbox{\c{c}}\else \c{c}\fi{}ois},
  journal = {Phys. Rev. Lett.},
  volume = {115},
  issue = {25},
  pages = {256803},
  numpages = {5},
  year = {2015},
  month = {Dec},
  publisher = {American Physical Society},
  doi = {10.1103/PhysRevLett.115.256803},
  url = {https://link.aps.org/doi/10.1103/PhysRevLett.115.256803}
}

@article{Prethermal_Without_T,
  title = {Prethermalization without Temperature},
  author = {Luitz, David J. and Moessner, Roderich and Sondhi, S. L. and Khemani, Vedika},
  journal = {Phys. Rev. X},
  volume = {10},
  issue = {2},
  pages = {021046},
  numpages = {11},
  year = {2020},
  month = {May},
  publisher = {American Physical Society},
  doi = {10.1103/PhysRevX.10.021046},
  url = {https://link.aps.org/doi/10.1103/PhysRevX.10.021046}
}

@article{Aarya,
author={Bothra, A. and Garg, A.},
title={Discrete Time Crystals in quantum Sherrington Kirkpatrick Model},
journal={arXiv},
volume={2504.19378},
year={2025},
}

@article{DTC1,
  title = {Observation of a discrete time crystal},
  author = {Zhang, J and Hess, P W and Kyprianidis, A and Becker, P and Lee, A and Smith, J and Pagano, G and Potirniche, I D and Potter, A C and Vishwanath, A and Yao, N Y and Monroe, C},
  journal = {Nature},
  volume = {543},
  issue = {},
  pages = {217–220},
  numpages = {4},
  year = {2017},
  month = {Mar},
  publisher = {Nature},
  doi = {10.1038/nature21413},
  url = {}
}

@article{DTC2,
  title = {Observation of discrete time-crystalline order in a disordered dipolar many-body system},
  author = {Choi, S and Choi, J and Landig, R and Kucsko, G and Zhou, H and Isoya, J and Jelezko, F and Onoda, S and Sumiya, H and Khemani, V and von Keyserlingk, C and Yao, N Y and Demler, E and Lukin, M D},
  journal = {Nature},
  volume = {543},
  issue = {},
  pages = {221–225},
  numpages = {5},
  year = {2017},
  month = {Mar},
  publisher = {Nature},
  doi = {10.1038/nature21426},
}

@article{DTC3,
  title = {Observation of Discrete-Time-Crystal Signatures in an Ordered Dipolar Many-Body System},
  author = {Rovny, Jared and Blum, Robert L. and Barrett, Sean E.},
  journal = {Phys. Rev. Lett.},
  volume = {120},
  issue = {18},
  pages = {180603},
  numpages = {5},
  year = {2018},
  month = {May},
  publisher = {American Physical Society},
  doi = {10.1103/PhysRevLett.120.180603},
  url = {https://link.aps.org/doi/10.1103/PhysRevLett.120.180603}
}

@article{Mori_2022,
  title = {Heating Rates under Fast Periodic Driving beyond Linear Response},
  author = {Mori, Takashi},
  journal = {Phys. Rev. Lett.},
  volume = {128},
  issue = {5},
  pages = {050604},
  numpages = {6},
  year = {2022},
  month = {Feb},
  publisher = {American Physical Society},
  doi = {10.1103/PhysRevLett.128.050604},
  url = {https://link.aps.org/doi/10.1103/PhysRevLett.128.050604}
}

@article{Bloch_Fermi_Hubbard,
  title = {Probing Transport and Slow Relaxation in the Mass-Imbalanced Fermi-Hubbard Model},
  author = {Darkwah Oppong, N. and Pasqualetti, G. and Bettermann, O. and Zechmann, P. and Knap, M. and Bloch, I. and F\"olling, S.},
  journal = {Phys. Rev. X},
  volume = {12},
  issue = {3},
  pages = {031026},
  numpages = {8},
  year = {2022},
  month = {Aug},
  publisher = {American Physical Society},
  doi = {10.1103/PhysRevX.12.031026},
  url = {https://link.aps.org/doi/10.1103/PhysRevX.12.031026}
}

@misc{Google_Fermi_Hubbard,
      title={Observation of separated dynamics of charge and spin in the Fermi-Hubbard model}, 
      author={Frank Arute and Arya et al, K.}, 
      year={2020},
      eprint={2010.07965},
      archivePrefix={arXiv},
      primaryClass={quant-ph},
      url={https://arxiv.org/abs/2010.07965}, 
}

@Article{Quantum_Dot_array_Fermi_Hubbard,
author={Hensgens, T.
and Fujita, T.
and Janssen, L.
and Li, Xiao
and Van Diepen, C. J.
and Reichl, C.
and Wegscheider, W.
and Das Sarma, S.
and Vandersypen, L. M. K.},
title={Quantum simulation of a Fermi--Hubbard model using a semiconductor quantum dot array},
journal={Nature},
year={2017},
month={Aug},
day={01},
volume={548},
number={7665},
pages={70-73},
issn={1476-4687},
doi={10.1038/nature23022},
url={https://doi.org/10.1038/nature23022}
}

@article{Anwesha1,
  title = {Gutzwiller projection for exclusion of holes: Application to strongly correlated ionic Hubbard model and binary alloys},
  author = {Chattopadhyay, Anwesha and Garg, Arti},
  journal = {Phys. Rev. B},
  volume = {97},
  issue = {24},
  pages = {245114},
  numpages = {18},
  year = {2018},
  month = {Jun},
  publisher = {American Physical Society},
  doi = {10.1103/PhysRevB.97.245114},
  url = {https://link.aps.org/doi/10.1103/PhysRevB.97.245114},
}

@article{TN2,
  title = {Accurate Simulation of the Hubbard Model with Finite Fermionic Projected Entangled Pair States},
  author = {Liu, Wen-Yuan and Zhai, Huanchen and Peng, Ruojing and Gu, Zheng-Cheng and Chan, Garnet Kin-Lic},
  journal = {Phys. Rev. Lett.},
  volume = {134},
  issue = {25},
  pages = {256502},
  numpages = {8},
  year = {2025},
  month = {Jun},
  publisher = {American Physical Society},
  doi = {10.1103/r4q9-4yvj},
  url = {https://link.aps.org/doi/10.1103/r4q9-4yvj}
}

@article{TN1,
title = {A practical introduction to tensor networks: Matrix product states and projected entangled pair states},
journal = {Annals of Physics},
volume = {349},
pages = {117-158},
year = {2014},
issn = {0003-4916},
doi = {https://doi.org/10.1016/j.aop.2014.06.013},
url = {https://www.sciencedirect.com/science/article/pii/S0003491614001596},
author = {Román Orús}
}

@article{Anwesha2,
  title = {Phase diagram of the half-filled ionic Hubbard model in the limit of strong correlations},
  author = {Chattopadhyay, Anwesha and Bag, Soumen and Krishnamurthy, H. R. and Garg, Arti},
  journal = {Phys. Rev. B},
  volume = {99},
  issue = {15},
  pages = {155127},
  numpages = {12},
  year = {2019},
  month = {Apr},
  publisher = {American Physical Society},
  doi = {10.1103/PhysRevB.99.155127},
  url = {https://link.aps.org/doi/10.1103/PhysRevB.99.155127}
}

@article{Mag_order,
author={Mentink, J. and Balzer, K. and Eckstein, M.},
title={Ultrafast and reversible control of the exchange interaction in Mott insulators.},
journal={Nat Commun.},
volume={6}, 
page={6708},
year={2015},
doi={https://doi.org/10.1038/ncomms7708},
}

@article{Anwesha3,
title={Unconventional superconductivity in a strongly correlated band-insulator without doping},
author={Chattopadhyay, A. and Krishnamurthy, H. R. and Garg, A.},
journal={SciPost Phys. Core},
volume={4},
pages={009},
year={2021},
doi={doi: 10.21468/SciPostPhysCore.4.2.009},
url={https://www.scipost.org/10.21468/SciPostPhysCore.4.2.009?acad_field_slug=astronomy},
}

@article{SM,
author={},
title={Supplemental Materials},
doi={},
url={},
}

@article{Bednorz,
  author={Bednorz, J. G. and M\''{u}ller, K. A.},
  title={Possible high Tc Superconductivity in the Ba- La-Cu-O System.},
  journal={Z. Phys. B- Condensed Matter},
  volume={64}, 
  page=189,
  year={1986}
  }

@article{pnictide_expt,
  author={Kamihara, Y. and Watanabe, T. and Hirano, M. and Hosono, H.},
  title={Iron-based layered superconductor $La[O_{1-x}F_x]FeAs (x = 0.05-0.12)$ with $T_c = 26 K$},
  journal={J. Am. Chem. Soc.},
  volume={130}, 
  page={3296},
  year={2008}
  }

@article{organic,
  author={Lefebvre et al, S.},
  title={Mott transition, antiferromagnetism, and unconventional superconductivity in layered organic superconductors},
  journal={Phys. Rev. Lett.},
volume={85}, 
page={5420},
year={2000}
}

@article{MTBLG,
  author={Cao et al, Y.},
  title={Unconventional superconductivity in magic-angle graphene superlattices},
  journal={Nature},
  volume={556}, 
  page={43}, 
  year={2018},
  }

@article{MTBLG2,
  author={Codecido et al, E.},
  title={Correlated insulating and superconducting states in twisted bilayer graphene below the magic angle},
  journal={Science Advances}, 
  volume={5},
  page={9}, 
  year={2019},
  }

@article{Lee,
     author={Lee, P. A. and Nagaosa, N. and Wen,  X. G.},
      title={Doping a Mott insulator: Physics of high-temperature superconductivity.},
      journal={Rev. Mod. Phys.},
      volume={78}, 
      page={17-85},
      year={2006},
      }

@article{Pnictides,
  author={Si, Q. and Yu, R. and Abrahams, E.},
  title={High temperature superconductivity in Iron Pnictides and Chalcogenides.},
  journal={Nature Rev. Mater.},
  volume={1}, 
  pages={16017},
  year={2016},
  }

@article{garg,
author={Garg, A. and Trivedi, N. and Randeria, M.},
title={Strong Correlations make high Tc superconductors robust against disorder},
journal={Nature Physics},
volume={4},
page={762},
year={2008},
doi={https://doi.org/10.1038/nphys1026},
url={https://www.nature.com/articles/nphys1026},
}

@article{Pan,
Author={Pan, S. H. and O'Neal, J. P. and Badzey, R. L. and Chamon, C. and Ding, H. and Engelbrecht, J. R. and Wang, Z. and Eisaki, H. and Uchida, S. and Gupta, A. K. and Ng, K.-W. and Hudson, E. W. and Lang, K. M. and Davis, J. C.},
title={Microscopic electronic inhomogeneity in the high-$T_c$ superconductor $Bi_2Sr_2CaCu_2O_{8+\delta}$.},
journal={Nature},
volume={413}, 
page={282-285},
year={2001},
}

@article{Mcelroy,
author={McElroy, K. and Lee, J. and Slezak, J. A. and  Lee, D. H. and Eisaki, H. and Uchida, S. and and Davis, J. C. },
title={Atomic-scale sources and mechanism of nanoscale electronic disorder in $Bi_2Sr_2CaCu_2O_{8+\delta}$.},
journal={Science},
volume={309}, 
paper={1048-1052},
year={2005},
}

@article{edegger-advphysics,
author={Edegger, B. and Gros, C. and Muthukumar, V.N.},
title={Gutzwiller-RVB theory of high temperature superconductivity: results from renormalised mean field theory and 
variational Monte Carlo calculations.},
journal={Adv. Phys.},
volume={56}, 
pages={927},
year={2007},
}

@article{Ogata,
  author={Ogata, M. and Himeda, A. },
  title={Superconductivity and antiferromagnetism in an extended gutzwiller approximation for $t–J$ Model: Effect of double-occupancy exclusion. },
  journal={J. Phys. Soc. Jpn.},
  volume={72}, 
  page={2}, 
  year={2003},
  }

@article{zhang,
author={Zhang, F. C. and Gros, C. and Rice, T. M. and Shiba, H.}, 
title={A renormalised hamiltonian approach for a resonant valence bond wavefunction.},
journal={Supercond. Sci. Tech.},
volume={1}, 
page={36},
year={1988},
}

@article{Arun,
author={Paramekanti, A. and Randeria, M. and Trivedi, N.}, 
title={Projected Wavefunctions and high temperature superconductivity.},
journal={Phys. Rev. Lett.},
volume={87},
page={217002}, 
year={2001},
}

@article{Cano,
  title = {Gate-tunable topological phases in superlattice modulated bilayer graphene},
  author = {Zeng, Yongxin and Wolf, Tobias M. R. and Huang, Chunli and Wei, Nemin and Ghorashi, Sayed Ali Akbar and MacDonald, Allan H. and Cano, Jennifer},
  journal = {Phys. Rev. B},
  volume = {109},
  issue = {19},
  pages = {195406},
  numpages = {9},
  year = {2024},
  month = {May},
  publisher = {American Physical Society},
  doi = {10.1103/PhysRevB.109.195406},
  url = {https://link.aps.org/doi/10.1103/PhysRevB.109.195406},
}

@article{Esslinger1,
  title = {Floquet Dynamics in Driven Fermi-Hubbard Systems},
  author = {Messer, Michael and Sandholzer, Kilian and G\"org, Frederik and Minguzzi, Joaqu\'{\i}n and Desbuquois, R\'emi and Esslinger, Tilman},
  journal = {Phys. Rev. Lett.},
  volume = {121},
  issue = {23},
  pages = {233603},
  numpages = {6},
  year = {2018},
  month = {Dec},
  publisher = {American Physical Society},
  doi = {10.1103/PhysRevLett.121.233603},
  url = {https://link.aps.org/doi/10.1103/PhysRevLett.121.233603}
}

@article{Esslinger2,
title={Enhancement and sign change of magnetic correlations in a driven quantum many-body system},
author={Görg, F. and Messer, M. and Sandholzer et al, K.},
journal={Nature},
volume={553},
pages={481},
year={2018},
doi={10.1038/nature25135},
url={https://doi.org/10.1038/nature25135}
}

@article{Dunlap,
  title = {Dynamic localization of a charged particle moving under the influence of an electric field},
  author = {Dunlap, D. H. and Kenkre, V. M.},
  journal = {Phys. Rev. B},
  volume = {34},
  issue = {6},
  pages = {3625--3633},
  numpages = {0},
  year = {1986},
  month = {Sep},
  publisher = {American Physical Society},
  doi = {10.1103/PhysRevB.34.3625},
  url = {https://link.aps.org/doi/10.1103/PhysRevB.34.3625}
}

@article{Holthaus,
  title = {Collapse of minibands in far-infrared irradiated superlattices},
  author = {Holthaus, Martin},
  journal = {Phys. Rev. Lett.},
  volume = {69},
  issue = {2},
  pages = {351--354},
  numpages = {0},
  year = {1992},
  month = {Jul},
  publisher = {American Physical Society},
  doi = {10.1103/PhysRevLett.69.351},
  url = {https://link.aps.org/doi/10.1103/PhysRevLett.69.351}
}

@article{Eckardt,
  title = {Superfluid-Insulator Transition in a Periodically Driven Optical Lattice},
  author = {Eckardt, Andr\'e and Weiss, Christoph and Holthaus, Martin},
  journal = {Phys. Rev. Lett.},
  volume = {95},
  issue = {26},
  pages = {260404},
  numpages = {4},
  year = {2005},
  month = {Dec},
  publisher = {American Physical Society},
  doi = {10.1103/PhysRevLett.95.260404},
  url = {https://link.aps.org/doi/10.1103/PhysRevLett.95.260404}
}

@article{Tsuji,
  title = {Dynamical Band Flipping in Fermionic Lattice Systems: An ac-Field-Driven Change of the Interaction from Repulsive to Attractive},
  author = {Tsuji, Naoto and Oka, Takashi and Werner, Philipp and Aoki, Hideo},
  journal = {Phys. Rev. Lett.},
  volume = {106},
  issue = {23},
  pages = {236401},
  numpages = {4},
  year = {2011},
  month = {Jun},
  publisher = {American Physical Society},
  doi = {10.1103/PhysRevLett.106.236401},
  url = {https://link.aps.org/doi/10.1103/PhysRevLett.106.236401}
}

@article{Watanabe,
author={Watanabe, T. and Ishihara, S.},
title={Band and Mott Insulators and Superconductivity in Honeycomb-
Lattice Ionic-Hubbard Model.},
journal={J. Phys. Soc. Jpn.},
volume={82},
pages={034704},
year={2013},
}

@article{Bloch2008,
  title = {Many-body physics with ultracold gases},
  author = {Bloch, Immanuel and Dalibard, Jean and Zwerger, Wilhelm},
  journal = {Rev. Mod. Phys.},
  volume = {80},
  pages = {885--964},
  year = {2008},
  doi = {10.1103/RevModPhys.80.885}
}
\end{document}


\renewcommand{\ni}{{\noindent}}
\newcommand{\dprime}{{\prime\prime}}
\newcommand{\be}{\begin{equation}}
\newcommand{\ee}{\end{equation}}
\newcommand{\bea}{\begin{eqnarray}} 
\newcommand{\eea}{\end{eqnarray}}
\newcommand{\la}{\langle}
\newcommand{\ra}{\rangle} 
\newcommand{\dg}{\dagger}
\newcommand\lbs{\left[}
\newcommand\rbs{\right]}
\newcommand\lbr{\left(}
\newcommand\rbr{\right)}
\newcommand\f{\frac}
\newcommand\e{\epsilon}
\newcommand\ua{\uparrow}
\newcommand\da{\downarrow}
\newcommand\mbf{\mathbf}

\title{Supplementary material for ``Periodic drive induced superconductivity in particle-hole symmetric systems"}
\maketitle
\begin{center}
    \textbf{The Floquet Hamiltonian of our model}
\end{center}
We obtain the Floquet Hamiltonian up to order $1/\Omega$ in the Magnus expansion $H_F=H_F^0+H_F^{[1]}$ 

\begin{equation}
\begin{split}
H_F^{[0]} = -t_0 \sum_{\langle ij \rangle, \sigma} \left( \mathcal{J}_0(\frac{A}{\Omega}sin(\Phi)) c_{i,\sigma}^\dagger c_{j,\sigma} + h.c \right) + U \sum_j n_{j \uparrow} n_{j \downarrow}+V_{0}\sum(n_{i\sigma}^{A}-n_{j\sigma}^{B}) -\mu\sum_i n_i, \\
H_F^{[1]}  = \frac{t_0^2}{4\pi i\hbar \Omega} \int_0^{2\pi} dt_1 \int_0^{t_1} dt_2 \sum_{\langle ij\rangle,\sigma} 
\left[
\big( g(t_1) g^*(t_2) - g^*(t_1) g(t_2) \big) 
\big( n_{i \sigma}^A - n_{j, \sigma}^B \big)
\right] \\
+ \frac{2t_0^2}{4 \pi i\hbar \Omega} \int_0^{2\pi} dt_1 \int_0^{t_1} dt_2 \sum_{(ij),\sigma}
\left[
\big( g(t_1) g^*(t_2) - g^*(t_1) g(t_2) \big) 
\big( c_{iA,\sigma}^\dagger c_{jA,\sigma}
- c_{iB, \sigma}^\dagger c_{jB, \sigma} \big) 
\right] +h.c.\\
+ \frac{t_0^2}{4 \pi i\hbar \Omega} \int_0^{2\pi} dt_1 \int_0^{t_1} dt_2 \sum_{((ij)),\sigma}
\left[
\big( g(t_1) g^*(t_2) - g^*(t_1) g(t_2) \big) 
\big( c_{iA,\sigma}^\dagger c_{jA,\sigma}
- c_{iB, \sigma}^\dagger c_{jB, \sigma} \big) 
\right] +h.c.\\
+ \frac{t_0U}{4\pi i\hbar \Omega} \int_0^{2\pi} dt_1 \int_0^{t_1} dt_2 \sum_{\langle ij\rangle}^z \sum_\sigma 
\left[
\big( g(t_1) - g(t_2)\big) \big(n_{i \sigma}^A -n_{j, \sigma}^B \big)c_{iA \bar{\sigma}}^\dagger c_{jB, \bar{\sigma}} + h.c.
\right]\\
+ \frac{t_0 V_{0}}{2\pi \iota \hbar \Omega} \int dt_1 \int dt_2 \sum_{i \in A, j=1}^z \sum_\sigma 
\left[
\big( g(t_1) - g(t_2) \big) 
\big( C_{i A\sigma}^{\dagger} C_{i+j,B \sigma}+h.c \big)
\right].
\end{split}
\end{equation}

In the above expansion $g(t)=\exp(i\frac{A}{\Omega}\sin(\Omega t)\sin(\Phi))$. While in $H_F^{[1]}$ the first term corresponds to a staggered potential between the two sub-lattices with the strength 
\be
V_{ind} = \frac{4t^2}{4\pi i\hbar \Omega} \int_0^{2\pi} dt_1 \int_0^{t_1} dt_2
\left[g(t_1) g^*(t_2) - g^*(t_1) g(t_2) \right]
  \ee
  
The second and third terms represent the drive-induced 2nd neighbour and 3rd neighbour intra-sub-lattice hopping given by $2t^\prime_{ind}$ and $t^\prime_{ind}$ respectively, where 
\be
t^\prime_{ind} = \frac{t^2}{4\pi i\hbar \Omega} \int_0^{2\pi} dt_1 \int_0^{t_1} dt_2
\left[g(t_1) g^*(t_2) - g^*(t_1) g(t_2) \right] = \frac{V_{ind}}{4}.
\ee
The fourth term corresponds to the correlated hopping term 
\be
t_{cor}= \frac{t_0U}{4\pi i\hbar \Omega} \int_0^{2\pi} dt_1 \int_0^{t_1} dt_2 \big( g(t_1) - g(t_2)\big).
\ee
The last term corresponds to an induced nearest neighbour hopping $t_{ind}$ with amplitude 
\be
t_{ind}= \frac{t V_{0}}{2\pi \iota \hbar \Omega} \int dt_1 \int dt_2 \big( g(t_1) - g(t_2) \big) .
\ee
In the main text, effective nearest neighbour hopping is defined as $t_{eff}=-t_0\mathcal{J}_0(\frac{A}{\Omega}sin(\Phi))+t_{ind}$. Here, $\mathcal{J}_l(\frac{A}{\Omega}\sin(\Phi))$ is the $l^{th}$ Bessel function of the first kind.

Effective couplings defined in $H_{eff}$ are all measured in units of $t_{eff}$, that is $U_{eff}=U/t_{eff}$, $V_{eff}=(V_0+V_{ind})/t_{eff}$, $t^\prime_{eff}=t^\prime_{ind}/t_{eff}$ and $t_{cor}^{eff}=t_{cor}/t_{eff}$. These effective couplings get enhanced as the system approaches the dynamical localization point.  

\noindent
    \textbf{Details of Gutzwiller approximation and spin asymmetric renormalized Bogoliubov-deGennes Approach :}
We solve the Floquet Hamiltonian in the limit $U_{eff}+2V_{eff} >> 1,t^\prime_{eff}$ and $2V_{eff} \gg 1$. In this limit doublon states on A sublattice and hole states on the B sublattice become energetically unfavorable. We therefore project out high-energy configurations and derive an effective low-energy Hamiltonian using a similarity (Schrieffer--Wolff) transformation on this Hamiltonian as derived in~\cite{Anwesha2}. 
This results in an effective Hamiltonian $H_{eff}$ as in Eq.[3] of the main draft which is defined in the projected Hilbert space which consists of states $|\Psi \rangle  = P| \Psi_{0} \rangle $  where the projection operator P eliminates doublons from sublattice A and holes from sublattice B in the unprojected state  $|\Phi_{0} \rangle$. 
We use here the Gutzwiller approximation to handle the projection as mentioned in the main draft. 
We then obtain the Gutzwiller renormalized Hamiltonian in which all operators are replaced by the corresponding operators in the unprojected space rescaled by the above mentioned Jastro factors based on Gutzwiller approximation. 

\begin{widetext}
\begin{equation}
\begin{split}
\mathcal{H}_{eff}/t_{eff}=-g_{t}\sum_{\langle ij \rangle ,\sigma}c_{iA\sigma}^{\dagger}c_{jB\sigma}+h.c.+2t^\prime_{eff}\sum_{(ij)}\big[g_{tA\sigma}c_{iA\sigma}^{\dagger}c_{jA\sigma}-g_{tB\sigma}c_{iB\sigma}^{\dagger}c_{jB\sigma}\big] \\+t^\prime_{eff}\sum_{((ij))}\big[g_{A\sigma}c_{iA\sigma}^{\dagger}c_{jA\sigma}-g_{B\sigma}c_{iB\sigma}^{\dagger}c_{jB\sigma}\big]
+Jg_{sAB}\sum_{\langle ij \rangle} \big[S_{iA}.S_{jB}-\frac{n_A(2-n_B)}{4}\big]\\+
4J^\prime g_{sA} \sum_{(ij)} \big[S_{iA}.S_{jA}-\frac{n_{iA}n_{jA}}{4}\big]+4J^\prime\sum_{(ij)}g_{sB} \big[S_{iB}.S_{jB}-\frac{(2-n_{iB})(2-n_{jB})}{4}\big]\\
+J^\prime g_{sA}\sum_{((ij))} \big[S_{iA}.S_{jA}-\frac{n_{iA}n_{jA}}{4}\big]+J^\prime\sum_{((ij))}g_{sB} \big[S_{iB}.S_{jB}-\frac{(2-n_{iB})(2-n_{jB})}{4}\big]\\
+ t_{\mathrm{cor}}^{eff}g_{t} \sum_{\sigma,<ij>}
\big[(n_{iA\bar{\sigma}} - n_{j B \bar{\sigma}})
c^{\dagger}_{iA\sigma}c_{j B\sigma}\big] +H_0+H_d-\mu\sum_{i}n_{i}
\end{split}
\label{tJ}
\end{equation}
\end{widetext}

Terms in $H_{eff}$ like $H_0$ and $H_d$ which have only density operators don't get renormalized under Gutzwiller approximation.

We finally solve the renormalized effective Hamiltonian as in Eq.~\ref{tJ} within spin-asymmetric Bogoliubov-deGennes approach. 
For this all the interaction terms in Eq.~\ref{tJ} are decomposed into local bond-pairing amplitudes $\Delta_{ij}=\langle c^\dagger_{i\uparrow}c_{j\downarrow}\rangle/2$, Fock shifts  $\chi_{ij,\sigma}=\langle c^\dagger_{i\sigma}c_{j\sigma}\rangle$, magnetization $m_{\alpha}=\langle n_{\alpha\uparrow}-n_{\alpha\downarrow}\rangle$ and sublattice density difference $\delta=\langle n_{B}-n_{A}\rangle$. Here $\alpha={A,B}$ is the sublattice index and $\langle n_\alpha\rangle$ is the density of particles at sublattice $\alpha$. The effective spin-asymmetric BdG Hamiltonians is then diagonalized and the fields defined above are calculated self-consistently.

\section{Effective Couplings in the Floquet Hamiltonian for different values of $V_0$}
In Fig[4] of the main paper we have presented results for the d-wave pairing amplitude and staggered magnetization for various initial values of the staggered potential $V_0$ for a fixed set of drive parameters. Here we present the couplings and hoppings in the Floquet Hamiltonian for different values of $V_0$ which is crucial to understand the $V_0$ dependence of the d-wave superconducting phase and the transition into AFM MI phase. 
      \begin{figure}
  \begin{center}
    \includegraphics[width=4.9in,angle=0]{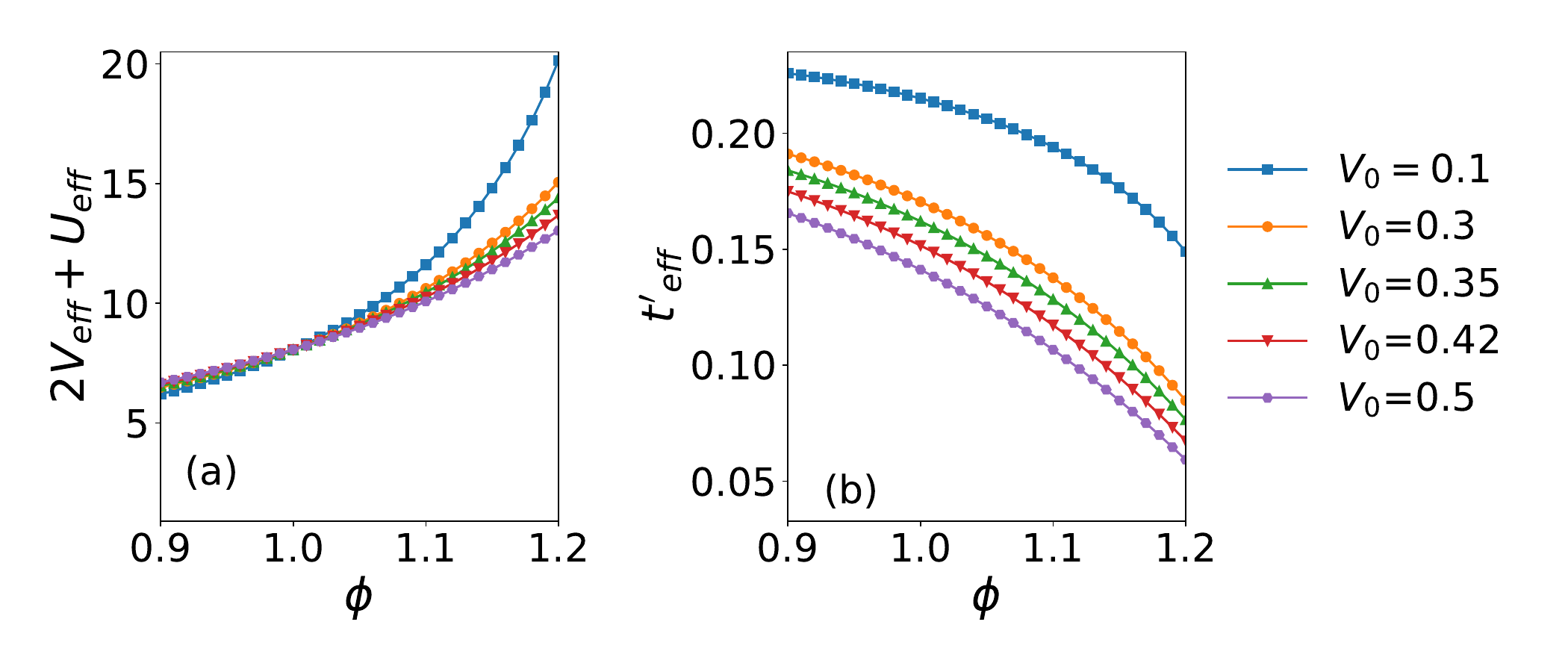}
  \includegraphics[width=4.5in,angle=0]{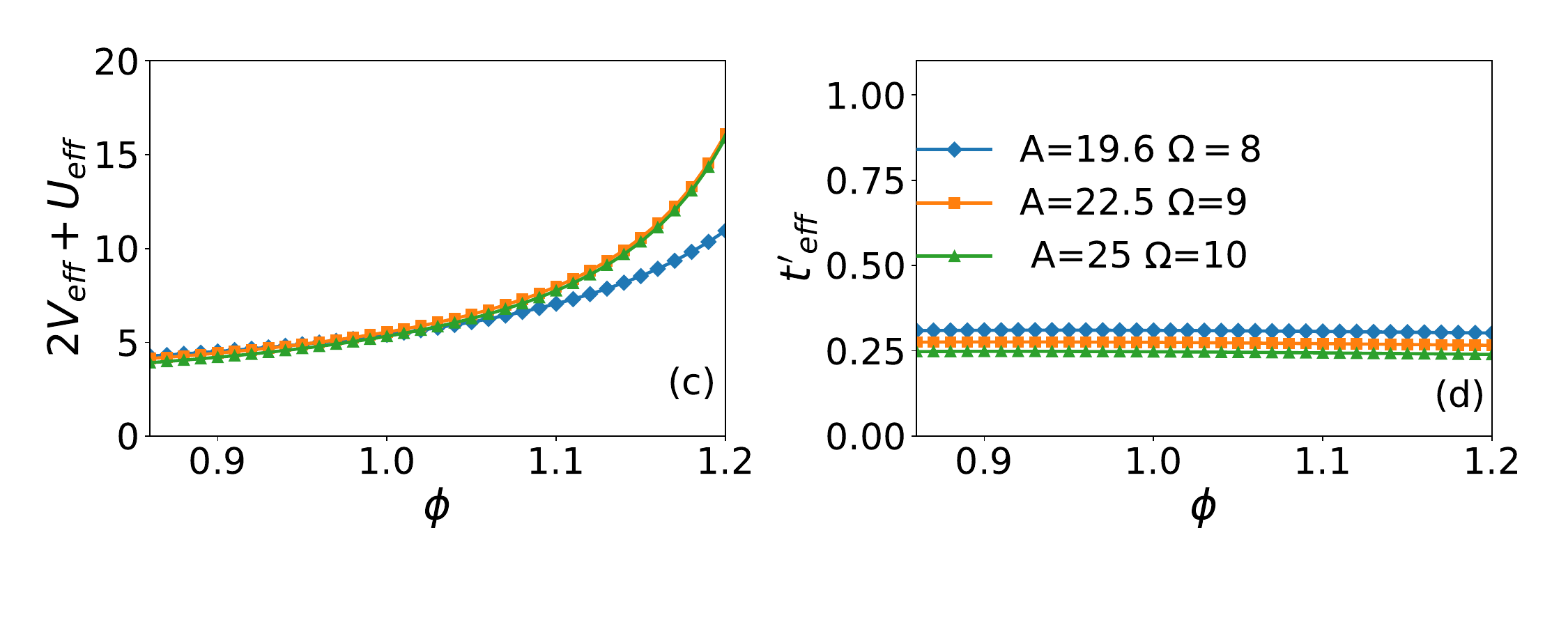}
          \caption{Panel[a] shows $U_{eff}+2V_{eff}$ vs $\Phi$ for various values of $V_0$ for $U=1.0t_0$. Panel[b] shows $t^\prime_{eff}$ vs $\Phi$ for the same parameters. The data shown is for the periodic drive with frequency $\Omega=10$ and amplitude $A=25$. Panel[c] and [d] presents the couplings $U_{eff}+2V_{eff}$ and $t^\prime_{eff}$ vs $\Phi$ for the system with $V_0=0$ and $U=0.55t_0$. Different curves correspond to different drive parameters as mentioned in the legend.}
          \label{couplings_V0}
          \end{center}
          \end{figure}

As shown in Fig.~\ref{couplings_V0}, as the strength of $V_0$ increases, up to $\Phi=1.0$ a slight increase in $U_{eff}+2V_{eff}$ occurs though for larger values of $\Phi$, the trend is reversed. But $t^\prime{eff}$ decreases monotonically as $V_0$ increases. This is because $V_0$ adds to effective nearest neighbour hopping through $t_{ind}$ which increases monotonically with $V_0$. As $t_{eff}$ increases, $t^\prime_{eff}=t^\prime_{ind}/t_{eff}$ decreases. Exactly because of the same reason $U_{eff}$ also decreases as $V_0$ increases. But $V_{eff}$ has two contributions $(V_0+V_{ind})/t_{eff}$ which lead to small enhancement for small $\Phi$ in $V_{eff}$. 
Though for the system to be in strongly correlated regime, $U_{eff}+2V_{eff} \gg 1$, but d-wave superconductivity can arise only if $J^\prime = 4(t^\prime_{eff})^2/U_{eff}$ is comparable to the nearest neighbour spin-exchange term $J=2/(U_{eff}+2V_{eff})$. As $V_0$ increases, $J$ increases though $J^\prime$ at some point starts decreasing. The regime in which $J^\prime \ge J$ AFM order is frustrated and the system goes into a d-wave superconducting phase. This explains why the width of the d-wave superconducting phase in $\Phi$ is larger for smaller values of $V_0$ as long as one can still be in the regime of strong correlation physics. 

Interestingly, d-wave superconductivity can be realized even for $V_0=0$ provided the periodic drive frequency and amplitude are in slightly lower range to give sufficiently large $t^\prime_{eff}$ and drive induced staggered magnetization $V_{ind}$. Panels[c] and [d] show the couplings and $t^\prime_{eff}$ for $V_0=0$ and $U=0.55t_0$ for three slightly lower frequency drives.Since $t^\prime_{eff}$ is larger here compared to the cases of higher frequency drives, frustration to AFM order can easily take place for a broad range of $\Phi$ resulting in a broad d-wave superconducting phase. 
\section{Couplings in the Floquet Hamiltonian for the alternative driving protocol}
\begin{figure}
  \begin{center}
    \includegraphics[width=4.9in,angle=0]{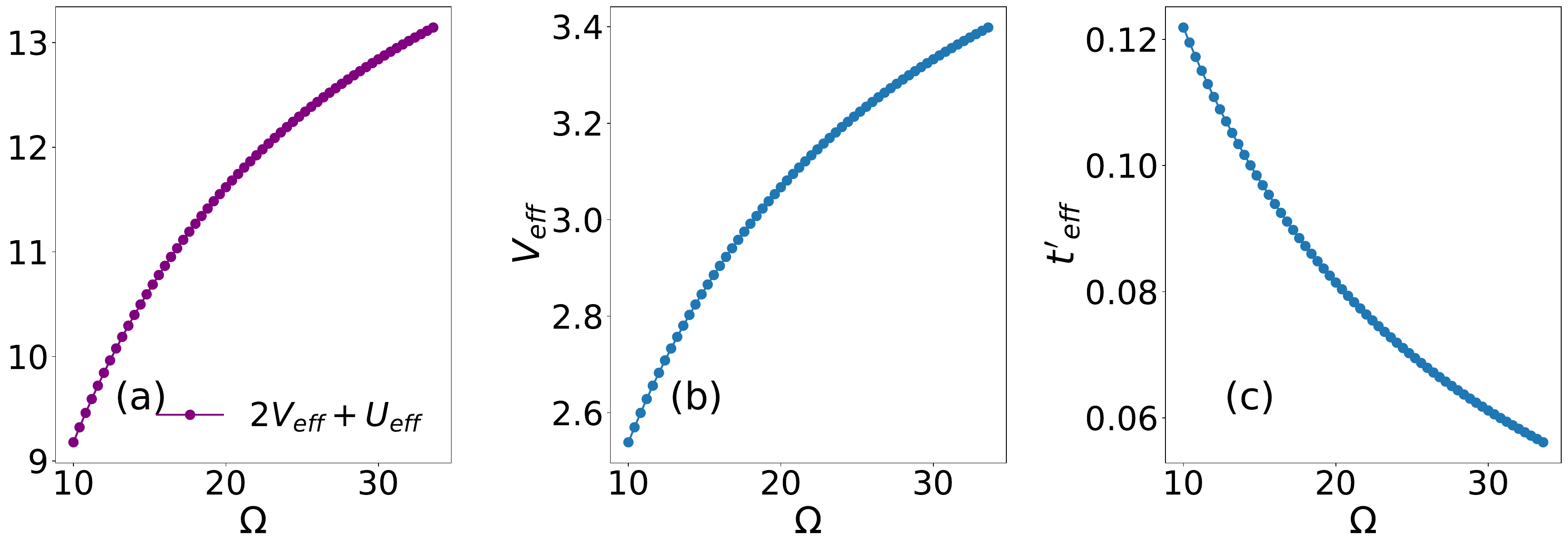}
          \caption{Panel[a] shows $U_{eff}+2V_{eff}$ vs $\Omega$ for $\Phi=1.07$. Panel[b] shows $t^\prime_{eff}$ vs $\Omega$ for the same parameters. The data shown is for the periodic drive with amplitude $A=2.5\Omega$, $U=1.0t_0$ and $V_0=0.5t_0$.}
          \label{couplings_phi}
          \end{center}
          \end{figure}

          The main draft presented the d-wave superconducting phase for alternative driving protocol in which the two sublattice drive has a fixed phase $\Phi$ and the drive frequency is tuned keeping $A/\Omega$ fixed. 
          Here, in Fig.~\ref{couplings_phi} we have shown $U_{eff}+2V_{eff}$ vs $\Omega$. Though the Bessel function part $\mathcal{J}_0\big(A/\Omega sin(\Phi)\big)$ in $t_{eff}$ does not change as a function of $\Omega$ for a fixed ratio of $A$ to $\Omega$, the drive induced part $t_{ind}$ decreases as the drive frequency increases resulting in overall decrease in $t_{eff}$ with increasing $\Omega$. This results in the enhancement of $U_{eff}+2V_{eff}$ as $\Omega$ increases which transfers the system into strongly correlated regime. But $t^\prime{eff}$ decreases, just like, $t_{ind}$ as $\Omega$ increases. Eventually at some large value of drive frequency the second neighbour spin-exchange $J^\prime$ can not frustrate the AFM order and the d-wave superconducting phase transits into the AFM MI.
          \section{Single-particle density of states in the AFM state}
\begin{figure}
  \begin{center}
    \includegraphics[width=4.9in,angle=0]{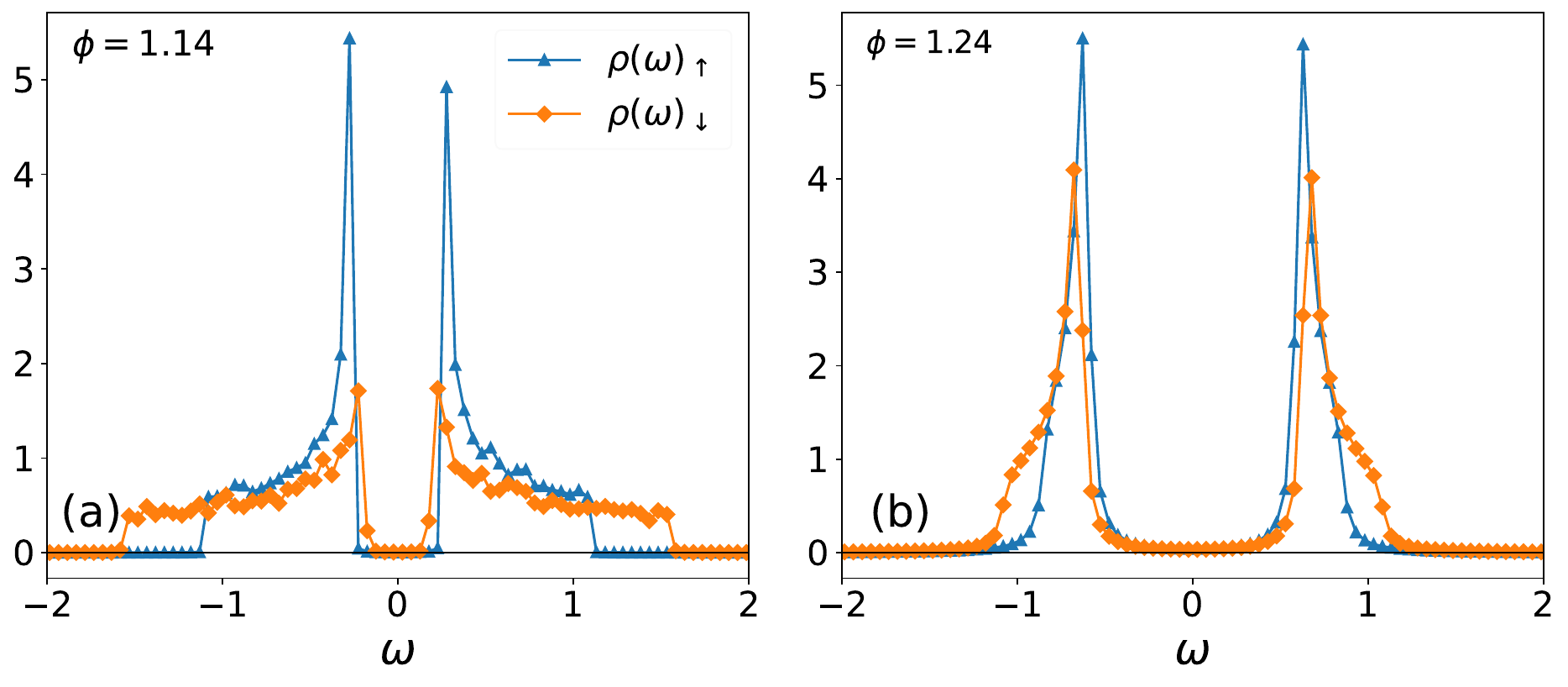}
          \caption{Single particle density of states for a few $\Phi$ values in the regime where the system has finite staggered magnetization. The data shown is for $\Omega=30$ and $=2.5\Omega$ with initial $U=1.0t_0$ and $V_0=0.5$}
          \label{dos}
          \end{center}
          \end{figure}

          As shown in the main paper, upon tuning the drive parameter (like phase $\Phi$) for a fixed drive frequency and amplitude  the system undergoes a transition from d-wave superconductor to AFM state with finite staggered magnetization. To confirm that this entire AFM phase is indeed a Mott Insulator, we calculated the single particle density of states for a few values of $\Phi$. As shown in Fig.~[\ref{dos}], the system indeed has a hard gap in the single particle excitation spectrum for the up and down spin density of states.

\bibliography{HM_drive.bib}